\documentclass[review]{elsarticle}

\usepackage{lineno,hyperref}
\usepackage{savesym}
\savesymbol{TRUE}
\usepackage{amsmath}
\usepackage{amsfonts}
\usepackage{mathrsfs}
\usepackage{graphicx}
\usepackage{epstopdf}
\usepackage{algorithm}
\usepackage{algorithmic}
\usepackage{color}
\usepackage{multirow}
\usepackage{subfigure}

\def\b0{\mathbf{0}}

\def\bb{\mathbf{b}}\def\bc{\mathbf{c}}

\def\bw{\mathbf{w}}
\def\bz{\mathbf{z}}

\def\bD{\mathbf{D}}
\def\bG{\mathbf{G}}\def\bH{\mathbf{H}}
\def\bI{\mathbf{I}}

\def\bQ{\mathbf{Q}}\def\bS{\mathbf{S}}
\def\bV{\mathbf{V}}\def\bW{\mathbf{W}}
\def\bY{\mathbf{Y}}\def\bZ{\mathbf{Z}}

\def\beps{{\boldsymbol{\epsilon}}}

\def\blambda{{\boldsymbol{\lambda}}}

\def\bphi{{\boldsymbol{\phi}}}
\def\bpsi{{\boldsymbol{\psi}}}

\def\bSigma{{\boldsymbol{\Sigma}}}
\def\btheta{{\boldsymbol{\theta}}}

\def\bxi{{\boldsymbol{\xi}}}

\biboptions{authoryear}

\journal{Computational Statistics \& Data Analysis}

\begin{document}

\begin{frontmatter}

\nolinenumbers

\title{Estimating Functional Linear Mixed-Effects Regression Models}

\author{Baisen Liu\fnref{label1}}
\author{Jiguo Cao\corref{cor1}\fnref{label2}}

\cortext[cor1]{Corresponding email: jiguo\_cao@sfu.ca}

\address[label1]{School of Statistics, Dongbei University of Finance and Economics, Dalian 116025, China}
\address[label2]{Department of Statistics and Actuarial Science, Simon Fraser University, Burnaby, BC V5A1S6, Canada}





\begin{abstract}
The functional linear model is a popular tool to investigate the relationship between a scalar/functional response variable and a scalar/functional covariate. We generalize this model to a functional linear mixed-effects model when repeated measurements are available on multiple subjects. Each subject has an individual intercept and slope function, while shares common population intercept and slope function. This model is flexible in the sense of allowing the slope random effects to change with the time. We propose a penalized spline smoothing method to estimate the population and random slope functions. A REML-based EM algorithm is developed to estimate the variance parameters for the random effects and the data noise. Simulation studies show that our estimation method provides an accurate estimate for the functional linear mixed-effects model with the finite samples. The functional linear mixed-effects model is demonstrated by investigating the effect of the 24-hour nitrogen dioxide on the daily maximum ozone concentrations and also studying the effect of the daily temperature on the annual precipitation.
\end{abstract}


\end{frontmatter}

\section{Introduction}

When a random variable is measured or observed at multiple time points or spatial locations, the data can be viewed as a function of time or spatial locations. This type of data is generally called  as functional data \citep{RamsaySilverman05}.  In the current big data era, functional data analysis (FDA) has become very popular in statistical methodology and applied data analysis. Functional linear models (FLMs) is one of the most popular models in FDA. It models the relationship between functional variables and/or predicts the scalar response from the functional input. FLMs have been studied extensively since \cite{RamsayDalzell91} introduced them.  With the developments of modern technology, FLMs have been popularly applied to model functional data in many fields such as economics, medicine, environment, climate [see for instance, \cite{RamsaySilverman02}, \cite{RamsaySilverman05}, and \cite{FerratyVieu06}, for several case studies].

There is extensive literature studying estimations and properties of FLMs. For example, \cite{ChiouMullerWang03} applied a quasi-likelihood
approach to study a FLM with a functional response and a finite-dimensional vector of scalar predictors. \cite{Yao2005} studied FLMs for sparse longitudinal data and suggested a nonparametric estimation method based on the functional principal components analysis (FPCA). Their proposed functional regression approach is flexible to allow for different measurement time points of functional predictors and the functional response.  \cite{CaiHall06} discussed the prediction problem in FLMs based on the FPCA technique. \cite{Crambes09} proposed a smoothing spline estimator for the functional slope parameter, and extended it to covariates with measurement-errors. \cite{YuanCai10} suggested a smoothness regularization method for estimating FLMs based on the reproducing kernel Hilbert space (RKHS) approach. They provided a unified treatment for both the prediction and
estimation problems by developing a tool on simultaneous diagonalization of two positive-definite kernels.  \cite{WuFan2010} proposed a varying-coefficient FLM which allows for the slope function depending on some additional scalar covariates. A systematic review on FLMs can be found in \cite{Morris15}.

One popular FLM is to link a scalar response variable $Y_j, j=1,\ldots,m,$ with a functional predictor $X_j(t)$ through the following model
\begin{equation}
\label{Eqn:FLM}
Y_j = \alpha+\int_\bold{\textit{S}} \beta(t) X_j(t) {\rm d}t + \epsilon_j,
\end{equation}
where $\alpha$ is the intercept, $\beta(t)$ is a smooth slope function, $\epsilon_j$'s are independent and identically distributed (i.i.d.) random variables with mean 0 and variance $\sigma^2_\epsilon$, and $\bold{\textit{S}}$ is often assumed to be a
compact subset of an Euclidean space such as $[0, 1]$. The slope function, $\beta(t)$, represents the accumulative effect of the functional covariate $X_j(t)$ on the scalar response $Y_j$.

For purposes of illustration, we take the air pollution data as an example. This data is from the R package NMMAPSdata \citep{Peng2004}.
The data have hourly measurements of ozone and nitrogen dioxide $NO_2$ concentrations for some U.S. cities.  Our aim is to study the relationship of the daily maximum ozone concentration and the functional predictors nitrogen dioxide $NO_2(t)$ measured during 24 hours (from 0 am to 11 pm) of that day. For the $i$-th city, the scalar response $Y_{ij}$ is the maximum ozone concentration during 24 hours (from 0 am to 11 pm) in the $j$-th day, and the functional covariate $X_{ij}(t)$ is the hourly $NO_2(t)$ concentration measured during 24 hours. In a preliminary analysis, we performed a functional linear regression model \citep{Cardot07} on each individual city and found that there was a dramatic variation of the estimated $\hat{\beta}(t)$. This indicates that each city has different effects of the hourly $NO_2(t)$ concentration on the daily maximum ozone. Therefore, it may not be appropriate to pool all the data of U.S. cities together and provide  only one average effect of hourly $NO_2(t)$ on the daily maximum ozone. On the other hand, we may not use all of the data information available if we fit the functional linear model for each individual city separately.

%
%

To address this dilemma, we generalize the FLM (\ref{Eqn:FLM}) to incorporate random effects into the slope function, and call it the functional linear mixed-effects model (FLMM). Assume that we repeatedly observed a distinct functional predictor and scalar outcome for each subject over several visits. Then the observed data has the structure $\{Y_{ij},X_{ij}(t)\}$ for $1\leq i \leq n$ and $1\leq j\leq m_i$, where $Y_{ij}$ is the $j$-th repeated measurement of the scalar response for the $i$-th subject, and $X_{ij}(t)$ is the corresponding functional predictor. The functional linear mixed-effects model can be expressed as following:
\begin{equation}
\label{Eqn:FLMR}
Y_{ij} = \alpha_0+a_i+\int_\bold{\textit{S}} [\beta(t)+b_i(t)] X_{ij}(t) {\rm d}t + \epsilon_{ij}\;,\;\; i=1,2,...,n, j=1,2,...,m_i,
\end{equation}
where $\alpha_0$ is the population intercept, $a_i$ is the intercept random effect, $\beta(t)$ represents the population effect of $X_{ij}(t)$ on $Y_{ij}$, $b_i(t)$ stands for the random effect of $X_{ij}(t)$ on $Y_{ij}$ for the $i$-th subject, and $\epsilon_{ij}$ is the i.i.d. random variable with mean 0 and variance $\sigma^2_\epsilon$. In this article, we assume that $a_i \sim N(0,\sigma^2_a)$, $\epsilon_{ij} \sim N(0,\sigma^2_\epsilon)$, and $b_i(t)$ follows a Gaussian stochastic process with mean 0 and covariance function $\gamma(s,t)$, that is, $b_i(t) \sim GP(0, \gamma(s,t))$. We also assume that $a_i$, $\epsilon_{ij}$, $b_i(t)$, and $X_{ij}(t)$  are mutually independent. The above functional linear mixed-effects model is very attractive, because it can estimate the population effect and random effect of the functional predictor $X(t)$(e.g. the hourly $NO_2(t)$ in the air pollution study) on the scalar response $Y$(e.g. daily maximum ozone concentrations) as well as the population intercept and intercept random effect simultaneously. The application of the proposed functional linear mixed-effects model on the air pollution problem is not unique, and many similar applications can be found in environmental or biological problems.

The proposed functional linear mixed-effect model (\ref{Eqn:FLMR}) is different from the following functional mixed model \citep{Goldsmith11, Goldsmith12}:
\begin{equation}\label{Eqn:Goldsmith}
Y_{ij} = \bZ_i\bb_i + \int_\bold{\textit{S}} \beta(t) X_{ij}(t) {\rm d}t + \epsilon_{ij},
\end{equation}
where $\bb_i \sim N(\b0,\sigma^2_\bb \bI)$ accounts for correlations in the repeated outcomes for the $i$-th subject. The highlight of the distinction between (\ref{Eqn:FLMR}) and (\ref{Eqn:Goldsmith}) is: the subject-specific random effect $\bb_i$ of (\ref{Eqn:Goldsmith}) remains the same across visits, while the random effect $b_i(t)$ of (\ref{Eqn:FLMR}) allows for varying with time. The including of the random effect $b_i(t)$ in (\ref{Eqn:FLMR}) can characterize the different trend effect of functional predictor on scalar outcomes for different subjects.

Many nonparametric smoothers used for the FLMs can be applied to fit the model (\ref{Eqn:FLMR}). In this article, we use the idea of penalized splines smoothers of \cite{RamsaySilverman05} to estimate $\beta(t)$ and $b_i(t)$ in (\ref{Eqn:FLMR}). Then, the model (\ref{Eqn:FLMR}) is transformed by a linear mixed-effects model (LMM). Then a REML-based EM algorithm is proposed to fit the LMMs, and its efficiency is illustrated by examples.

 The remainder of this
article is organized as follows. Section 2 introduces a smoothing spline method to estimate the above functional linear mixed-effects model.  Section 3 implemented some simulations to evaluate the finite sample performance of the smoothing spline method. Then the functional linear mixed-effects model is demonstrated by two real applications in Section 4.  Conclusions are given in Section 5.

\section{Method}
Without giving any parametric assumption on the slope functions, $\beta(t)$ and $b_i(t)$, we estimate them as linear combinations of splines basis functions
$$
\beta(t) = \sum_{j=1}^J c_j \phi_j(t) = \bphi'(t)\bc\;,~~b_i(t) = \sum_{k=1}^K b_{ik} \psi_k(t)= \bpsi'(t)\bb_i\;.
$$
where $\bphi(t)=(\phi_1(t),...,\phi_J(t))'$ and $\bpsi(t)=(\psi_1(t),...,\psi_K(t))'$ are two vectors of basis functions with dimensions $J$ and $K$, respectively, and $\bc=(c_1,...,c_J)'$ and $\bb_i=(b_{i1},...,b_{iK})'$ are the corresponding vectors of basis coefficients to estimate.
Let  $\bD = {\rm Cov}(\bb_i)={\rm E}(\bb_i\bb'_i)$ denote the variance-covariance matrix of random-effects, then $\bb_i \sim N(0,\bD)$, and the covariance function $\gamma(s,t)$ for the random effect $b_i(t)$ can be expressed as $\gamma(s,t) = \bpsi'(s)\bD\bpsi(t)\,.$ We first consider the scenario of the functional predictors $X_{ij}(t)$ observed without measurement errors. When $X_{ij}(t)$ is observed with measurement errors, many nonparametric smoothing approaches can be applied to reconstruct the underlying functional predictors $X_{ij}(t)$, such as the functional principal component analysis method \citep{Yao2005}, which is introduced in Subsection 2.5.

\subsection{Estimating Fixed and Random Effects}
Let $\btheta=(\alpha_0,\bc')'$ and $\bxi=(\bxi_1',...,\bxi_n')'$ with $\bxi_i=(a_i, \bb_i')', i=1,...,n$. The fixed effects $\{\alpha_0, \beta(t)\}$, and random effects $\{a_i, b_i(t)\}$ are estimated by minimizing
\begin{eqnarray}
\label{eqn:LogLIK}
H(\btheta,\bxi) &=& \sum\limits_{i=1}^n \sum\limits_{j=1}^{m_i}  \frac{1}{2\sigma^2_\epsilon}\bigg(Y_{ij} - \alpha_0 - a_i - \int_\bold{\textit{S}} [\beta(t)+b_i(t)] X_{ij}(t) {\rm d}t \bigg)^2+\frac{1}{2}\sum\limits_{i=1}^n\bb'_i\bD^{-1}\bb_i \nonumber\\
&& +  \bigg[\frac{\lambda_{\beta}}{2}\int_\bold{\textit{S}}  \bigg\{\frac{{\rm d}^2\beta(t)}{{\rm d}t^2}\bigg\}^2{\rm d}t + \frac{\lambda_{b}}{2}\sum_{i=1}^n \int_\bold{\textit{S}}  \bigg\{\frac{{\rm d}^2 b_i(t)}{{\rm d}t^2}\bigg\}^2{\rm d}t\bigg] + \frac{1}{2\sigma^2_a}\sum_{i=1}^n a_i^2\;.
\end{eqnarray}
Define three vectors $\bY_i = (Y_{i1},...,Y_{im_i})'$, $\bW_i=[\bw_{i1},\cdots,\bw_{im_i}]'$ and $\bZ_i=[\bz_{i1},\cdots,\bz_{im_i}]'$ where $ \bw_{ij} = \left(1, \int_\bold{\textit{S}} \bphi'(t)X_{ij}(t){\rm d}t\right)'$ and $\bz_{ij} = \left(1, \int_\bold{\textit{S}} \bpsi'(t)X_{ij}(t){\rm d}t\right)'$. Define a $J\times J$ matrix $\bG = \int_\bold{\textit{S}} ({\rm d}^2\bphi(t)/{\rm d}t^2)({\rm d}^2\bphi(t)/{\rm d}t^2)'{\rm d}t$ and a $K\times K$ matrix $\bG_b = \int_\bold{\textit{S}} ({\rm d}^2\bpsi(t)/{\rm d}t^2)({\rm d}^2\bpsi(t)/{\rm d}t^2)'{\rm d}t$. Then $H(\btheta,\bxi)$ can be expressed in a matrix form
\begin{eqnarray*}
H(\btheta,\bxi) &=& \sum\limits_{i=1}^n  \frac{1}{2\sigma^2_\epsilon}\|\bY_{i}-\bW_{i}\btheta-\bZ_{i}\bxi_i\|^2+\frac{1}{2}\sum\limits_{i=1}^n\bb'_i\bD^{-1}\bb_i \\
& & ~~~~~+ (\frac{\lambda_{\beta}}{2}\bc'\bG\bc + \frac{\lambda_b}{2}\sum\limits_{i=1}^n \bb'_i\bG_b\bb_i) +\frac{1}{2\sigma^2_a}\sum_{i=1}^n a_i^2\;.
\end{eqnarray*}
Then the estimates for $\btheta$ and $\bxi$ are obtained by minimizing $H(\btheta,\bxi)$:
\begin{equation}
\label{Eqn:PESc}
\begin{array}{ccl}
\hat{\btheta} &=& \left(\bW'\widetilde{\bV}^{-1}\bW+\lambda \widetilde{\bG}\right)^{-1} \bW'\widetilde{\bV}^{-1}\bY, \\
\hat{\bxi} &=& (\bI_n \bigotimes \widetilde{\bD}_\xi) \bZ'\widetilde{\bV}^{-1}(\bY-\bW\hat{\btheta})\;.
\end{array}
\end{equation}
where $\bY=(\bY'_1,...,\bY'_n)'$, $\bW=(\bW'_1,...,\bW'_n)'$, $\widetilde{\bV} = {\rm diag}(\tilde{\bV}_1,...,\tilde{\bV}_n)$
 with $\tilde{\bV}_i = \bZ_i \widetilde{\bD}_\xi \bZ'_i+\sigma^2_\epsilon\bI_{m_i}, i=1,...,n$,
 $\widetilde{\bG}={\rm diag}(0, \bG)$, $\widetilde{\bD}_\xi = (\bD^{-1}_\xi+\lambda_b \bG_\xi)^{-1}$, $\bD_\xi={\rm diag}(\sigma^2_a, \bD)$, $\bG_\xi={\rm diag}(0, \bD_b)$, $\bZ={\rm diag}(\bZ'_1,...,\bZ'_n)$, and $\bigotimes$ denotes the kronecker product.

Once obtaining the estimates $\hat{\btheta}=(\hat{\alpha}_0, \widehat{\bc}')'$ and $\hat{\bxi}_i=(\hat{a}_i, \widehat{\bb}_i')'$, the estimates of $\beta(t)$ and $b_i(t), i=1,...,n$, can be given by
\begin{equation}
\label{Eqn:PSmoothers}
\hat{\beta}(t) = \bphi'(t)\widehat{\bc},~~~\hat{b}_i(t) = \bpsi'(t)\widehat{\bb}_i, \;\;i=1,...,n.
\end{equation}

\subsection{The REML-based EM algorithm}

To estimate the fixed-effects, $\btheta$, the random-effects, $\bxi$, and the variance parameters, $\sigma^2_a$, $\sigma^2_\epsilon$ and $\bD$, we recommend an EM algorithm procedure called the \emph{REML-based EM-algorithm}. It was proposed by \cite{WuZhang2006} for estimating nonparametric mixed-effects regression models with longitudinal data. The REML-based EM-algorithm has three steps, which are outlined as follow.

{\bf Initializing}. Initializing the starting values for $\sigma^{2}_a$, $\sigma^{2}_\epsilon$ and $\bD$, denoted by $\sigma^{2(0)}_a$, $\sigma^{2(0)}_\epsilon$ and $\bD^{(0)}$\;, respectively. For example, we can choose $\sigma^{2(0)}_a = \sigma^{2(0)}_\epsilon = 1$ and $\bD^{(0)}$ as an identity matrix\;.

{\bf Step 1}. Set $r=r+1$. Compute
$$
\begin{array}{ccl}
\widetilde{\bD}^{(r-1)}_{\xi} &=& [\{\bD^{(r-1)}_{\xi}\}^{-1}+\lambda_b\bG_{\xi}]^{-1},\\
\tilde{\bV}^{(r-1)}_{i} &=& \bZ_i\widetilde{\bD}^{(r-1)}_{\xi}\bZ'_i+\sigma^{2(r-1)}_\epsilon\bI_{m_i}, i=1,2,...,n.
\end{array}
$$
Denote $\tilde{\bV}^{(r-1)} = {\rm diag}(\tilde{\bV}^{(r-1)}_1,...,\tilde{\bV}^{(r-1)}_n)$. Then estimate $\hat{\btheta}^{(r)}$ and $\hat{\bxi}^{(r)}_i$ by
$$
\begin{array}{ccl}
\hat{\btheta}^{(r)} &=& [\bW'\{\tilde{\bV}^{(r-1)}\}^{-1}\bW+\lambda_\beta \tilde{\bG}]^{-1}\bW'\{\tilde{\bV}^{(r-1)}\}^{-1}\bY,\\
\hat{\bxi}^{(r)}_i &=&\widetilde{\bD}^{(r-1)}_{\xi} \bZ'_i \{\tilde{\bV}^{(r-1)}_i\}^{-1}(\bY_i-\bW_i\hat{\btheta}^{(r)}), i=1,2,...,n.
\end{array}
$$

{\bf Step 2}. Compute the residuals $\hat{\beps}^{(r)}_i = \bY_i-\bW_i\hat{\btheta}^{(r)}-\bZ_i\hat{\bxi}^{(r)}_i$ and the matrix ${\bH}^{(r-1)}_i =\{\tilde{\bV}^{(r-1)}_i\}^{-1}-\{\tilde{\bV}^{(r-1)}_i\}^{-1}\bW_i[\bW'\{\tilde{\bV}^{(r-1)}\}^{-1}\bW+\lambda_\beta\tilde{\bG}]^{-1}
\bW'_i\{\tilde{\bV}^{(r-1)}_i\}^{-1}$. Then the updates of $\sigma^{2(r)}_{\epsilon}$ and $\bD^{(r)}_{\xi}={\rm diag}(\sigma^{2(r)}_a, \bD^{(r)})$ are given by
$$
\begin{array}{ccl}
\sigma^{2(r)}_\epsilon &=& N^{-1}\sum\limits_{i=1}^n\left\{\{\hat{\beps}^{(r)}_i\}'\hat{\beps}^{(r)}_i+\sigma^{2(r-1)}_\epsilon[m_i-\sigma^{2(r-1)}_\epsilon{\rm trace}({\bH}^{(r-1)}_i)]\right\},\\
\bD^{(r)}_\xi &=& n^{-1}\sum\limits_{i=1}^n \left\{\hat{\bxi}^{(r)}_i\{\hat{\bxi}^{(r)}_i\}' + [\bD^{(r-1)}_\xi-\bD^{(r-1)}_\xi\bZ'_i\bH^{(r-1)}_i\bZ_i\bD^{(r-1)}_\xi]\right\}.
\end{array}
$$

{\bf Step 3}. Repeat Steps 2 and 3, until some convergence conditions are satisfied.

\subsection{Smoothing Parameter Selection}
The smoothness of $\beta(t)$ and $b_i(t), i = 1, \ldots,n$, are controlled by the smoothing parameter $\lambda_{\beta}$ and $\lambda_b$, respectively. Define $\blambda = (\lambda_{\beta}, \lambda_b)'$, then the optimal value for $\blambda$ is chosen by minimizing the generalized cross-validation (GCV) criterion defined as follows
\begin{equation*}
\label{Eqn:GCV}
{\rm GCV}(\blambda) = \dfrac{\rm SSE(\blambda)}{(N-df(\blambda))^2},
\end{equation*}
where ${\rm SSE}(\blambda) = \sum\limits_{i=1}^n  \|\bY_i-\bW_i\widehat{\btheta}-\bZ_i\widehat{\bxi}_i\|^2$, $N=\sum_{i=1}^n m_i$, and $df(\blambda)$ is the effective degrees of freedom, which is calculated as $df(\blambda)={\rm trace}(\bQ)$, where  $\bQ$ is given by
\begin{equation*}
\label{Eqn:SmoothMatrix1}
\bQ = (\bW,\bZ)\left[ \dfrac{1}{\sigma^2_\epsilon}\left(
\begin{array}{cc}
\bW'\bW & \bW'\bZ\\
\bZ'\bW & \bZ'\bZ
\end{array}\right)+\left(
\begin{array}{cc}
\lambda_\beta\tilde{\bG} & 0\\
0 & \bI_n \bigotimes \widetilde{\bD}_\xi
\end{array}\right)
\right]^{-1}\left(
\begin{array}{c}
\bW'\\
\bZ'
\end{array}
\right).
\end{equation*}
 Define the matrix $\bS_i = \bW_i[\bW'\tilde{\bV}^{-1}\bW+\lambda_\beta\tilde{\bG}]^{-1}\bW'\tilde{\bV}^{-1}$. The predictor $\hat{\bY}_i = \bW_i\hat{\btheta}+\bZ_i \hat{\bxi}_i$ can then be expressed as $\hat{\bY}_i = \bQ_i\bY $ with
\begin{equation*}
\label{Eqn:SmoothMatrix2}
\bQ_i = \bS_i+\bZ_i\widetilde{\bD}_\xi\bZ'_i\tilde{\bV}^{-1}(\bI_{m_i}-\bS_i).
\end{equation*}
Note that the smooth matrix $\bQ=(\bQ'_1,...,\bQ'_n)'$.

\subsection{Constructing the Confidence Intervals}

To construct the confidence intervals of $\alpha_0$ and the point-wise confidence bands of $\beta(t)$, we need to calculate the covariance matrix of $\hat{\btheta}$:
\begin{eqnarray}
{\rm Cov}(\hat{\btheta}) &=& \left(\sum_{i=1}^n \bW'_i\tilde{\bV}^{-1}_i\bW_i+\lambda_\beta \widetilde{\bG}\right)^{-1}\left( \sum_{i=1}^n \bW'_i\tilde{\bV}^{-1}_i {\rm Cov}(\bY_i) \tilde{\bV}^{-1}_i \bW_i\right)  \nonumber\\
 & & \left(\sum_{i=1}^n \bW'_i\tilde{\bV}^{-1}_i\bW_i+\lambda_\beta \widetilde{\bG}\right)^{-1}.
\label{Eqn:CovC}
\end{eqnarray}
In (\ref{Eqn:CovC}), ${\rm Cov}(\bY_i)$ can be replaced by $\tilde{\bV}_i$ to account for our roughness penalty on $b_i(t), i=1,...,n$. For simplicity, instead of using  ${\rm Cov}(\hat{\btheta})$, we use
\begin{equation}
\label{Eqn:CovC2}
{\rm Cov}(\hat{\btheta}) = \left(\sum_{i=1}^n \bW'_i\tilde{\bV}^{-1}_i\bW_i+\lambda_\beta \widetilde{\bG}\right)^{-1}\;.
\end{equation}

Let $\widehat{\rm Cov}(\hat{\btheta})$ be the estimator of the covariance matrix (\ref{Eqn:CovC2}) and partition it as $\widehat{\rm Cov}(\hat{\btheta})=\left(
\begin{array}{cc}
\hat{\sigma}^2_{11} & \hat{\bSigma}_{12} \\
\hat{\bSigma}_{12}' & \hat{\bSigma}_{22}\\
\end{array}\right) $. Then the 95\% confidence intervals of $\alpha_0$ is approximately as
$$
(\hat{\alpha}_0-1.96\hat{\sigma}_{11}, \;\; \hat{\alpha}_0+1.96\hat{\sigma}_{11})\;,
$$
and the 95\% pointwise bands of $\beta(t)$ can be approximately given by
$$
\left(\hat{\beta}(t)-1.96\sqrt{\widehat{\rm Var}[\hat{\beta}(t)]}, \;\; \hat{\beta}(t)+1.96\sqrt{\widehat{\rm Var}[\hat{\beta}(t)]}\right), \mbox{~ for all ~} t \in \bold{\textit{S}},
$$
where $ \widehat{\rm Var}[\hat{\beta}(t)] = \Phi'(t) \hat{\bSigma}_{22}\Phi(t)$. Moreover, the estimate of $\gamma(s,t)$ can be given as
$$
\hat{\gamma}(s,t) = \Psi'(s) \widehat{\bD}_b \Psi(t)\;,
$$
where we use $\widehat{\bD}_b=(\b0, \bI_K)\left(\widehat{\bD}^{-1}_\xi+\lambda_b \bG_\xi\right)^{-1}(\b0, \bI_K)'$ instead of $\widehat{\bD}$ in order to account for our roughness penalty on $b_i(t),i=1,...,n$ in our method.

\subsection{Reconstructing the predictors $X_{ij}(t)$}
When covariates $X_{ij}(t)$ in the functional linear mixed-effects model (\ref{Eqn:FLMR}) are not be exactly observable but measured with errors, the estimators and inference may be biased if one ignores these measurement errors. Hence, we need to adjust the resulting bias. In this article, we suggest to reconstruct the functional predictors $X_{ij}(t)$ by using a large number of functional principal components
obtained from a smooth estimator of the covariance matrix estimator \citep{Goldsmith12} firstly. Then, we treat the estimated $\hat{X}_{ij}(t)$ as the true predictors and applying the REML-based EM algorithm.

Define the covariance function of $X(t)$ as
\begin{equation*}
\label{Eqn:Cov}
C(s,t)={\rm Cov}(X(t),X(s)).
\end{equation*}
Mercer's theorem \citep{Ash75} states that $C(s,t)$ has the eigen-decomposition
$$
C(s,t) = \sum\limits_{k=1}^\infty \lambda_k \varphi_k(s)\varphi_k(t),
$$
where $\lambda_1\geq \lambda_2\geq ... \geq 0$ satisfying $\sum_{k=1}^\infty \lambda_k <\infty$, and $\varphi_k(t)$'s form a complete orthonormal basis in
$\bold{\textit{S}} \times \bold{\textit{S}}$. Then, $X(t)$ allows the Karhunen-Loeve decomposition \citep{Rice1991}
$$
X(t)=\mu(t)+\sum\limits_{k=1}^\infty \xi_k\varphi_k(t)\;
$$
where $\varphi_k(\cdot)$ is the orthonormal eigenfunction, which is also called the functional principal component (FPC). The coefficients $\xi_k$ is called the FPC score of $X(t)$, which satisfies $E(\xi_k)=0$, $E(\xi_k^2)=\lambda_k$, and $E(\xi_k\xi_l)=0$ for $k\neq l$.

Suppose we have the following additive measurement error model,
$$
W_{ij}(t) = X_{ij}(t)+e_{ij}(t),
$$
where $W_{ij}(t)$ is the observed value, $X_{ij}(t)$ is the underlying true value for the $i$th
subject at the time point $t$, and $e_{ij}(t)$ represents the measurement error at the time point $t$. We assume that $e_{ij}(t)$ is a mean zero process, and $\{X_{ij}(t), e_{ij}(t)\}$ are mutually independent. We estimate $C(s,t)$ by using a method-of-moments approach, and then smooth
the off-diagonal elements of this observed covariance matrix to remove the `nugget effect' that
is caused by measurement error \citep{StaniswalisLee98, Yao2005, Goldsmith12}.

We use the principal analysis by conditional estimation (PACE) algorithm proposed by \cite{Yao2005} to estimate the mean curve $\mu_i(t)$, the FPCs $\varphi_{ik}(t)$ and the FPC scores $\xi_{ijk}$ from the  observations $W_{ij}(t_{ijk})$. Let $\hat{\mu}_i(t)$, $\hat{\varphi}_{ik}(t)$, and $\hat{\xi}_{ijk}$ be the corresponding estimators of $\mu_i(t)$,  $\varphi_{ik}(t)$ and $\xi_{ijk}$, respectively. Then an estimate of $X_{ij}(t)$ is obtained as
$$
\hat{X}_{ij}(t)= \hat{\mu}_i(t)+\sum\limits_{k=1}^M \hat{\xi}_{ijk}\hat{\varphi}_{ik}(t)\;
$$
where the number of FPCs, $M$, can be chosen by AIC, BIC, the cross-validation method, or the empirical experience based
on the percentage of explained variance (such as 90\% or 95\%).

\section{Simulation studies}

In this section, we perform some numerical experiments to assess the efficiency of our proposed estimating procedure for the functional linear mixed-effects model (\ref{Eqn:FLMR}). The performance of our estimation method is evaluated by the following relative mean integrated square error
(RMISE) for the estimated population slope function $\hat{\beta}(t)$ and the individual slope function $\hat{\beta}_1(t),\cdots,\hat{\beta}_n(t)$,
\begin{equation*}
\label{Eqn:RMISEbeta}
{\rm RMISE} (\hat{\beta}(t)) = \dfrac{\int_\bold{\textit{S}}(\hat{\beta}(t)-\beta(t))^2dt}{\int_\bold{\textit{S}} \beta^2(t)dt},
\end{equation*}
and
\begin{equation*}
\label{Eqn:RMISEsubj}
{\rm RMISE}(\hat{\beta}_1(t),\cdots,\hat{\beta}_n(t)) = \dfrac{\sum_{i=1}^n\int_\bold{\textit{S}}(\hat{\beta}_i(t)-\beta_i(t))^2dt}{\sum_{i=1}^n\int_\bold{\textit{S}} \beta^2_i(t)dt},
\end{equation*}
where $\beta_i(t) = \beta(t) + b_i(t)$, and $\hat{\beta}_i(t) = \hat{\beta}(t) + \hat{b}_i(t), i=1,\ldots,n$.

We assume that the functional predictor $X_{ij}(t)$ are observed at $n$ equally-spaced time points $\{t_k, k=1,...,n\}$  of $[0, 1]$  with the additive normal measurement errors:
$$
W_{ij}(t_k)=X_{ij}(t_k)+e_{ijk},~~ e_{ijk} \sim N(0, \sigma^2_e),
$$
where $\sigma_e=0.0$ or $0.5$, and the true underlying predictors $X_{ij}(t)$ are given by
$$
X_{ij}(t) = \mu_i(t) +\sqrt{2}\sum_{k=1}^4 \xi_{ijk}\psi_k(t)\;, ~ t\in [0,1]\;,
$$
where $\mu_i(t)=\delta_{i0}+\delta_{i1}\sin(\pi t)$ with independent random variables $\delta_{i0}\sim U[-2,2]$, $\delta_{i1} \sim N(0,4)$,  $\xi_{ijk} \sim N(0, 2/2^k), k=1,2,3,4$, and $\psi_1(t)=\sin(2\pi t)$, $\psi_2(t)=\cos(2\pi t)$, $\psi_3(t)= \sin(4\pi t)$ and $\psi_4(t)= \cos(4\pi t)$.
We choose two types of functions for the individual slope functions $\beta_i(t)$: (1) $\beta_i(t) = \eta_{0i}+\eta_{1i}t^2+\eta_{2i}\exp(-3t)$ with the population slope function given by $\beta(t) = 1+2t^2+\exp(-3t)$; (2) $\beta_i(t) = \eta_{0i}+\eta_{1i}\sin(2\pi t)+\eta_{2i}\cos(2\pi t)$ with the population slope function given by $\beta(t) = 1+2\sin(2\pi t)+\cos(2\pi t)$. The random coefficients are generated as $(\eta_{0i},\eta_{1i},\eta_{2i})' \sim N((1.0,2.0,1.0)'$, ${\rm diag}(0.2^2,0.4^2,0.2^2))$ for both of cases. The scalar response is generated from the following model:
$$
Y_{ij} = \alpha_i+\int_0^1 \beta_i(t) X_{ij}(t){\rm d}t + \epsilon_{ij}, \epsilon_{ij} \sim N(0,\sigma^2_\epsilon), \; i=1,2,...,n, j=1,2,...,m_i\;,
$$
where $\alpha_i$ is generated from $N(3.0,0.25)$. The number of repeated measurements for each individual is varied as $m_i=5,10,20$. The number of individuals is set as $n=50$ and $100$, and the standard deviation of the data noises is varied as $\sigma_\epsilon=0.5$ and $1.0$.

\begin{figure}
\begin{minipage}{0.48\linewidth}
\centerline{\includegraphics[width=6.0cm]{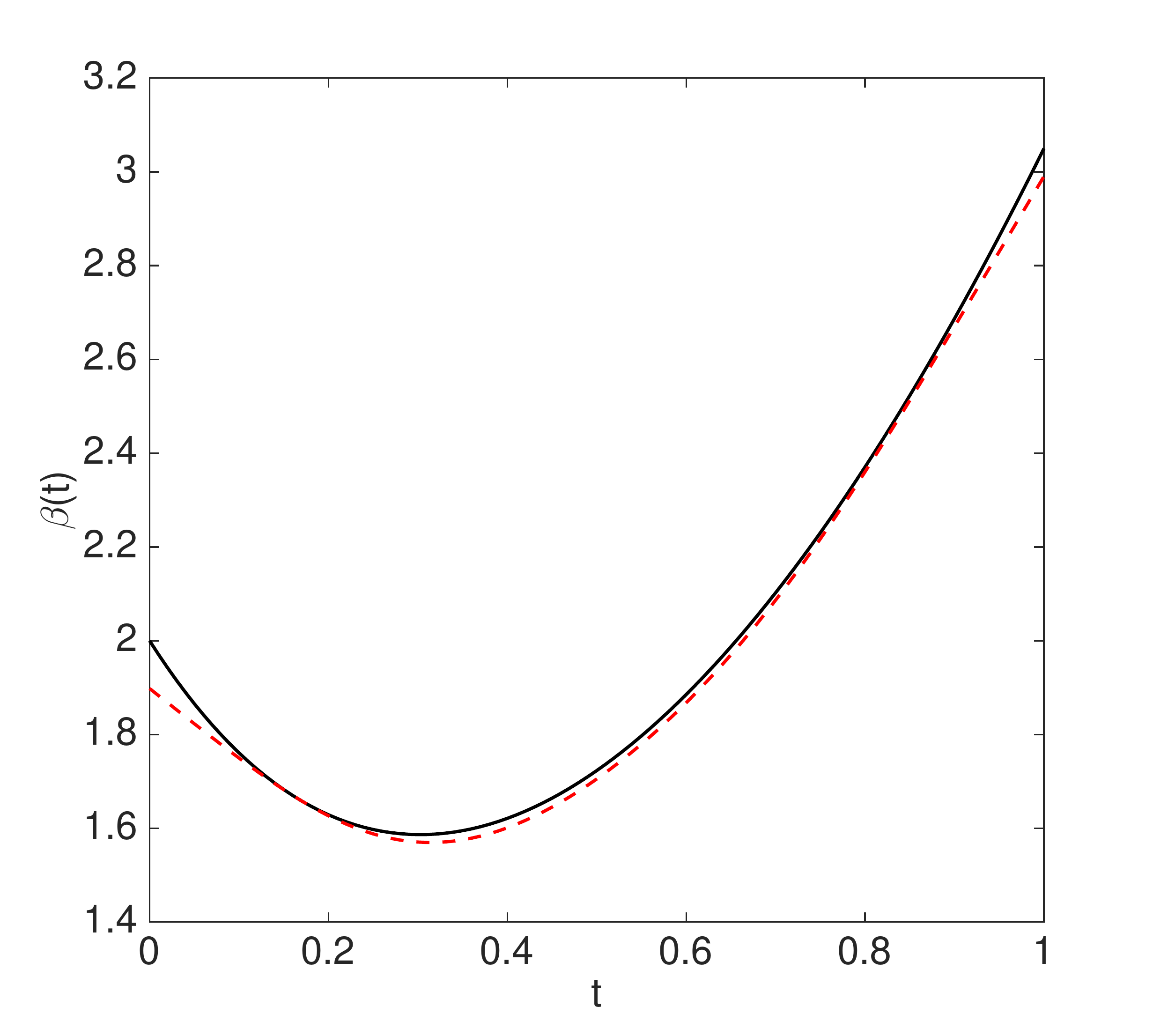}}
\end{minipage}
\hfill
\begin{minipage}{.48\linewidth}
\centerline{\includegraphics[width=6.0cm]{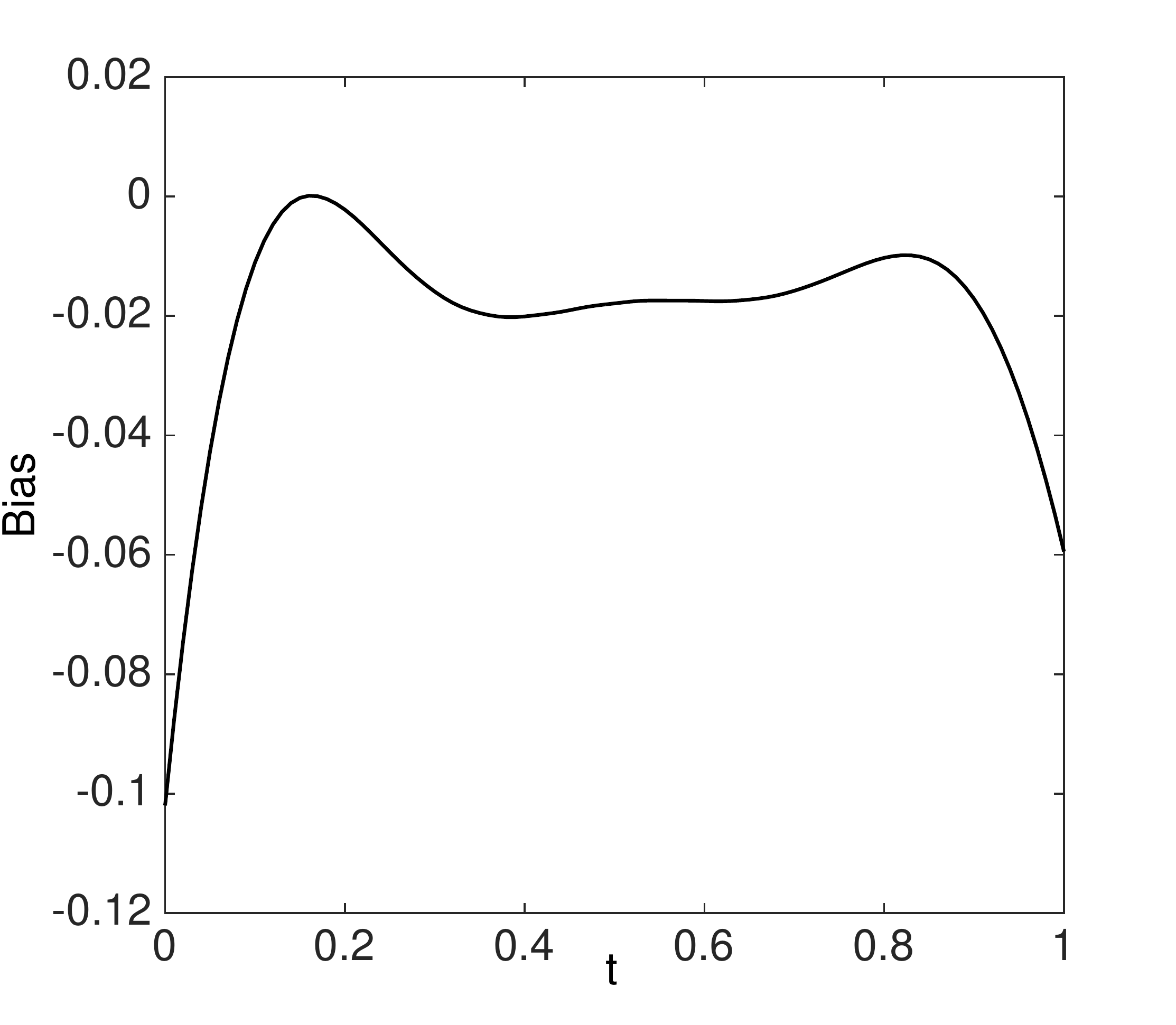}}
\end{minipage}
\vfill
\begin{minipage}{0.48\linewidth}
\centerline{\includegraphics[width=6.0cm]{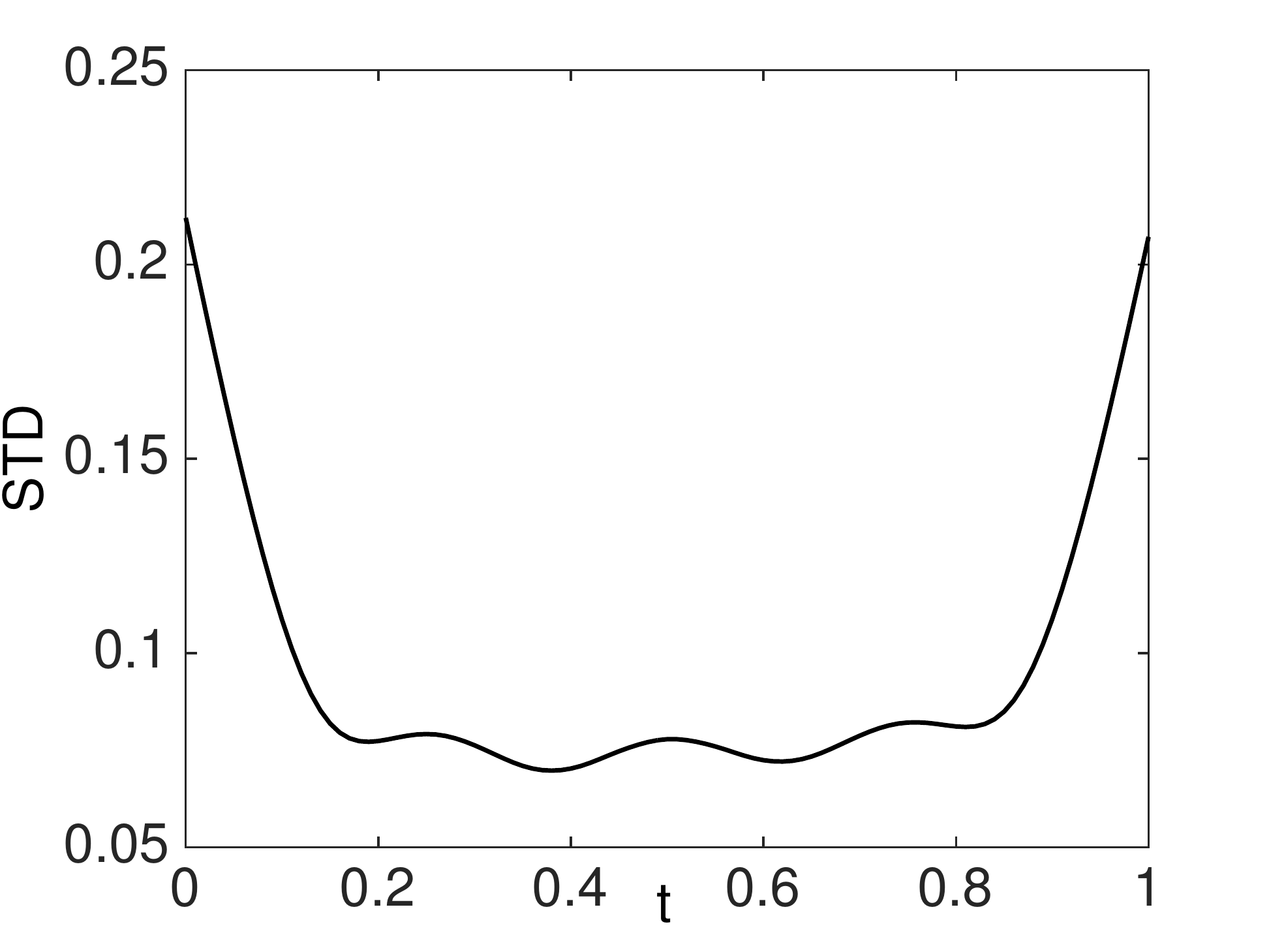}}
\end{minipage}
\hfill
\begin{minipage}{0.48\linewidth}
\centerline{\includegraphics[width=6.0cm]{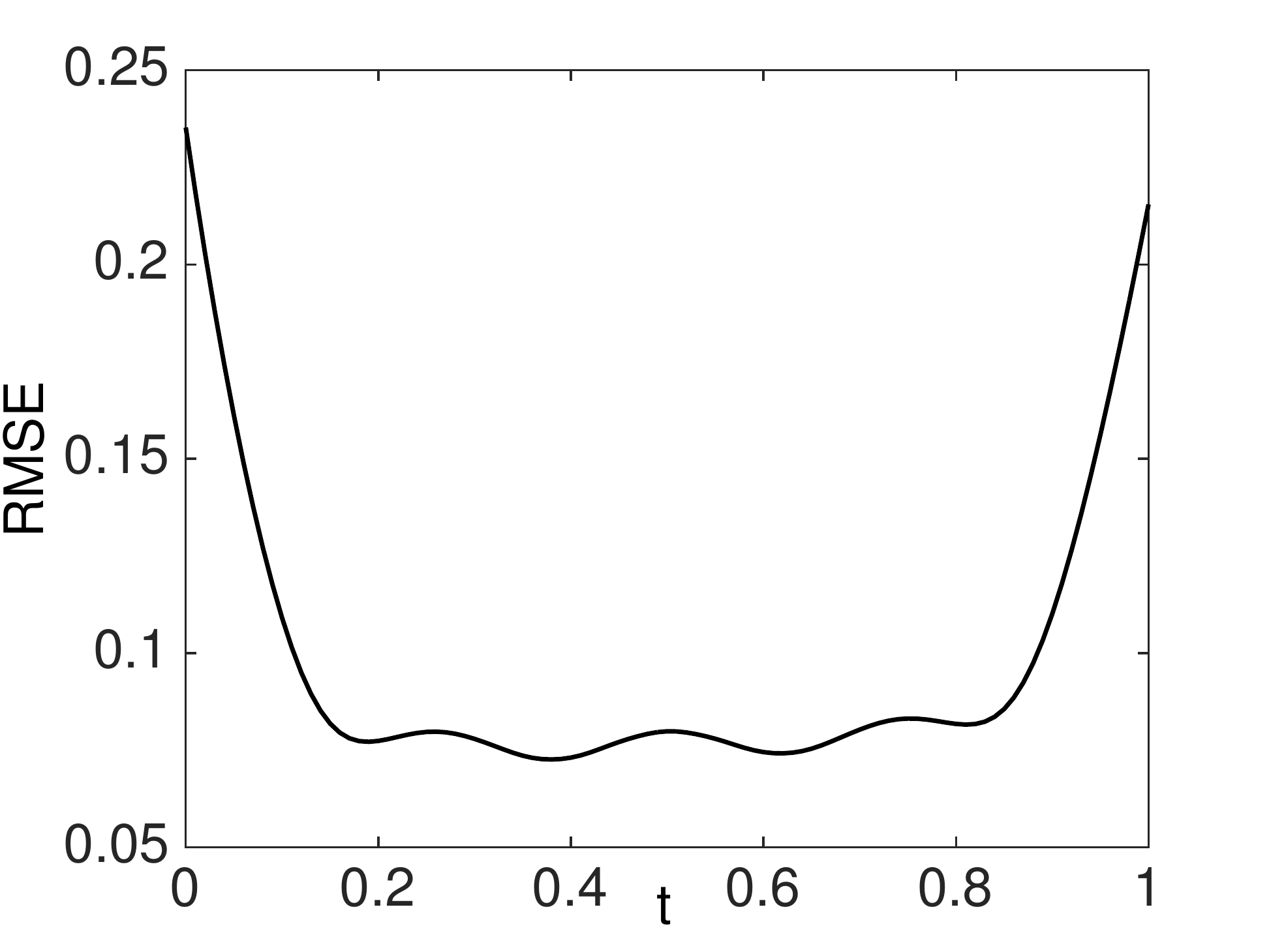}}
\end{minipage}
\caption{The pointwise mean, bias, standard deviation (STD) and root mean squared error (RMSE) of the estimated population slope function $\hat{\beta}(t)$ in 1,000 simulation replicates when $m_i=20$ and $\sigma_\epsilon=1.0$ in our simulation studies. The dashed line in the top left panel is the true  population slope function $\beta(t)=1+2t^2+\exp(-3t)$.}
\label{fig:Case1}
\end{figure}

\begin{figure}
\begin{minipage}{0.48\linewidth}
\centerline{\includegraphics[width=6.0cm]{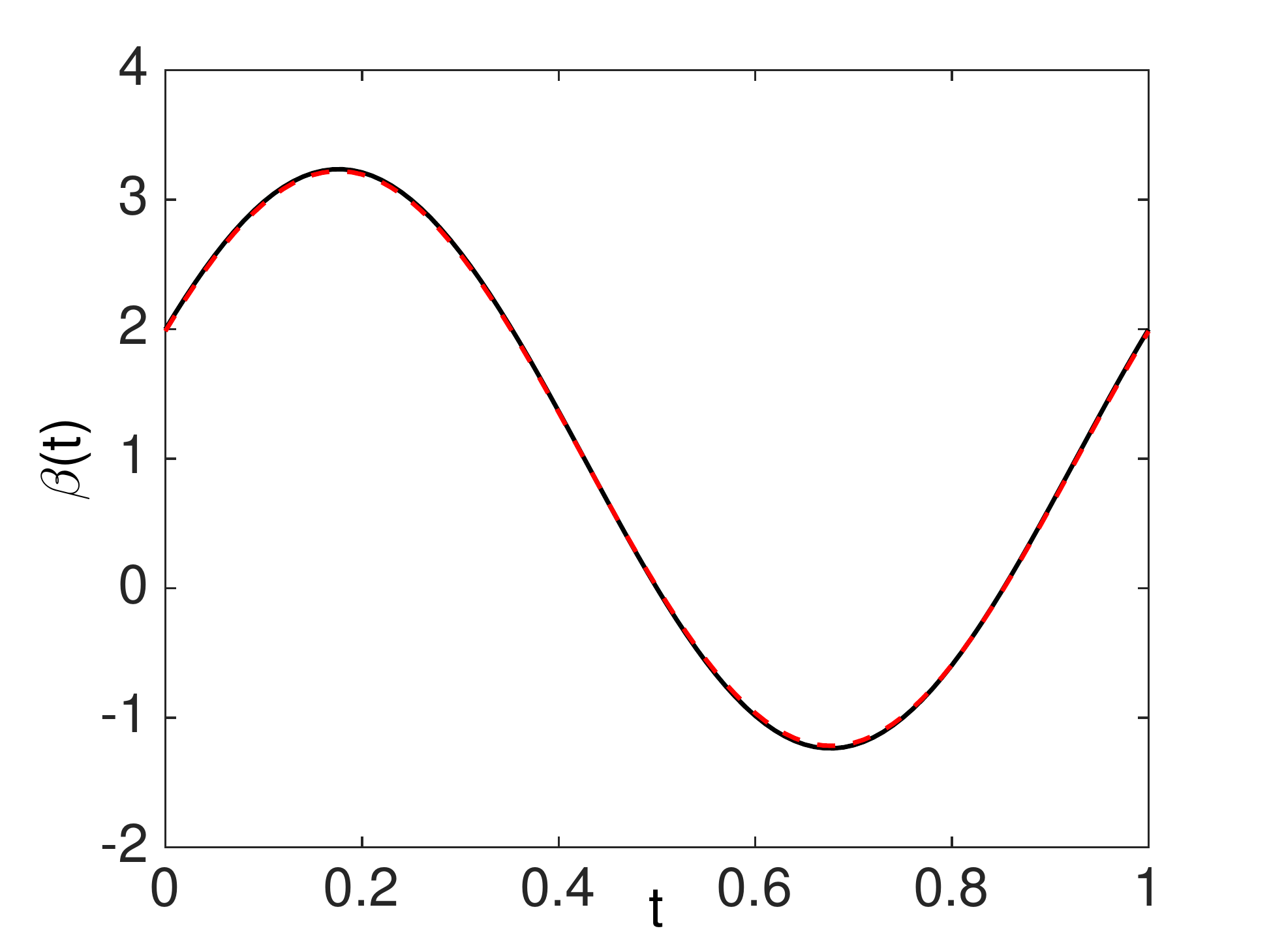}}
\end{minipage}
\hfill
\begin{minipage}{.48\linewidth}
\centerline{\includegraphics[width=6.0cm]{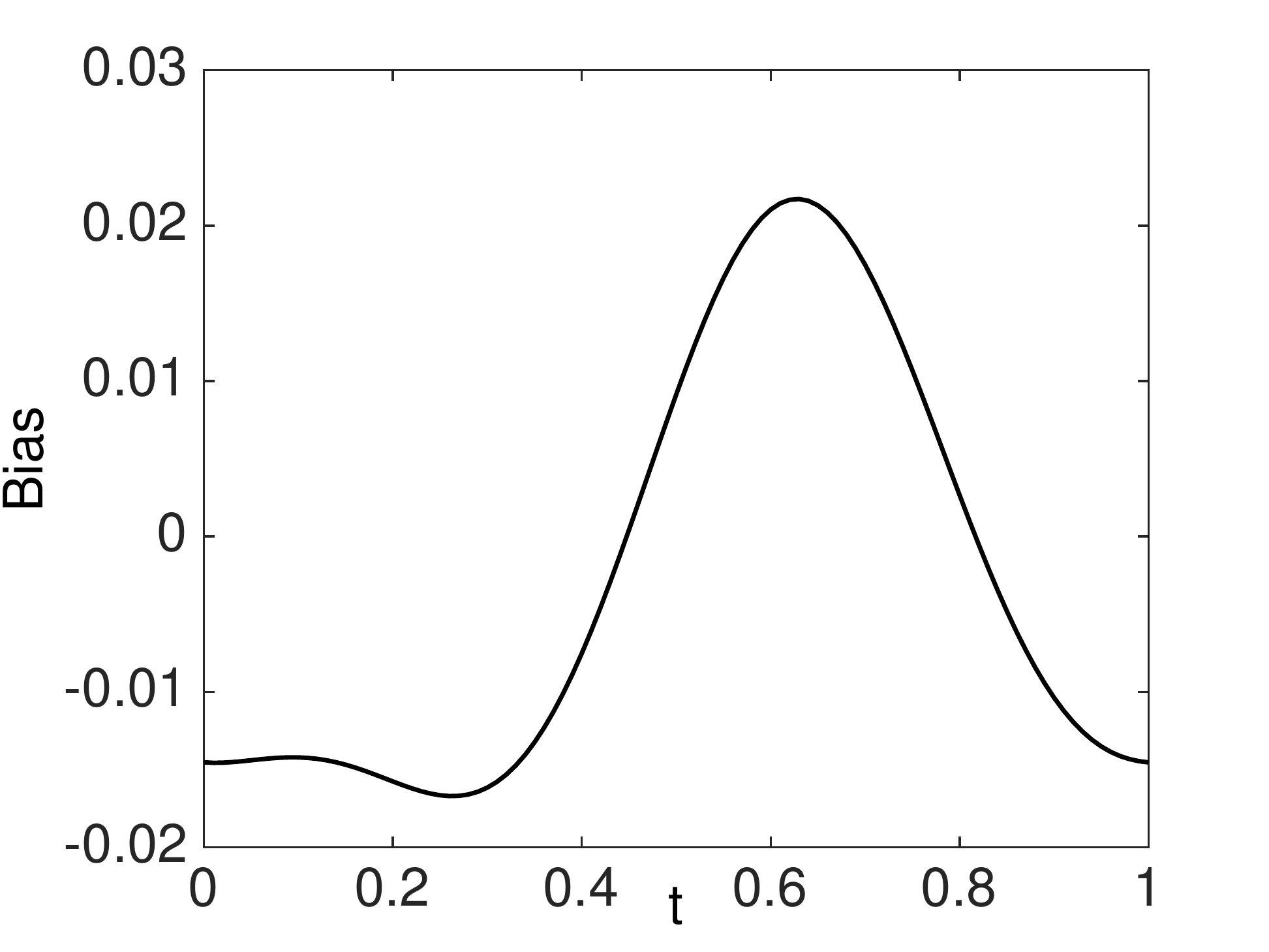}}
\end{minipage}
\vfill
\begin{minipage}{0.48\linewidth}
\centerline{\includegraphics[width=6.0cm]{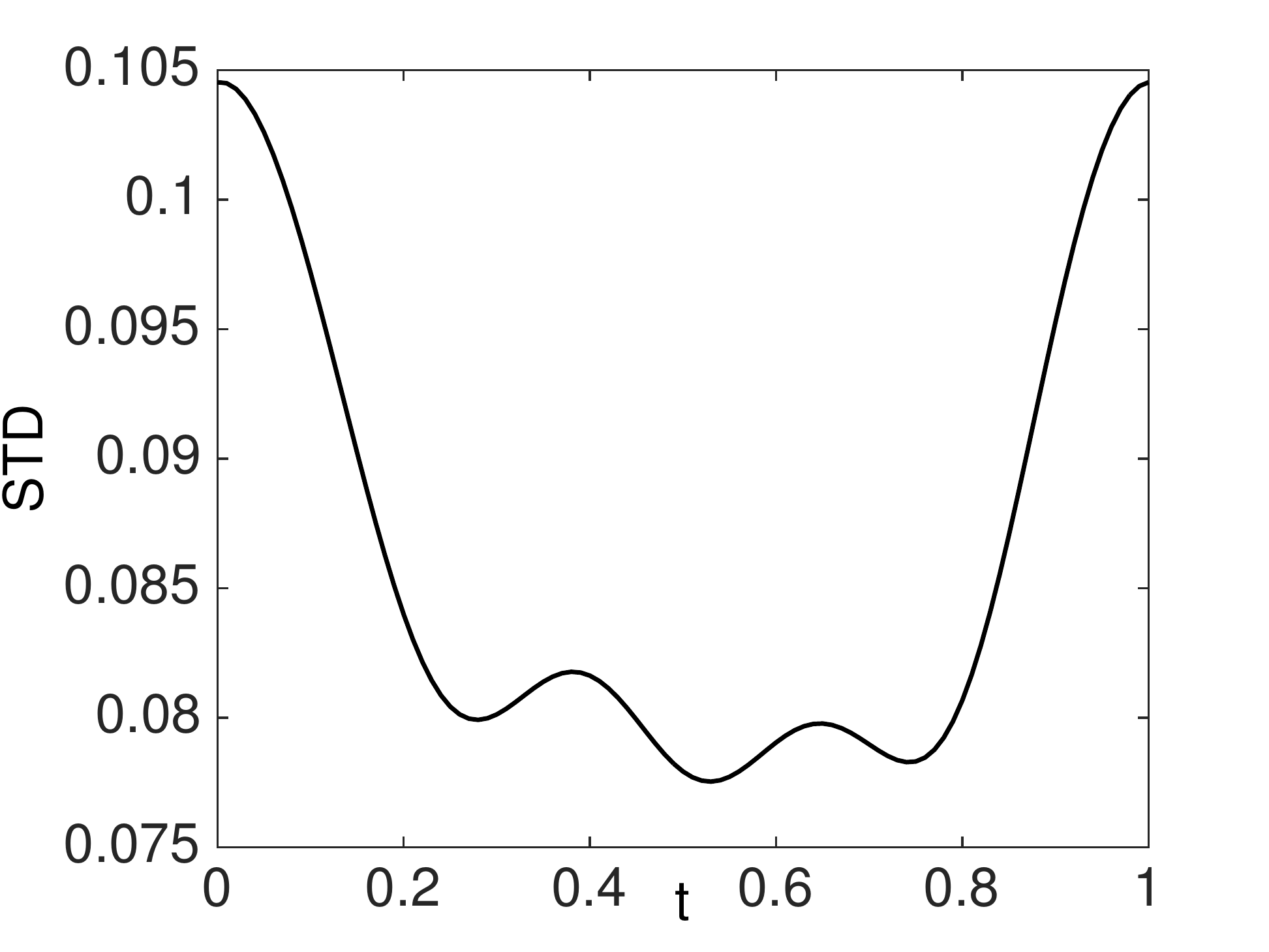}}
\end{minipage}
\hfill
\begin{minipage}{0.48\linewidth}
\centerline{\includegraphics[width=6.0cm]{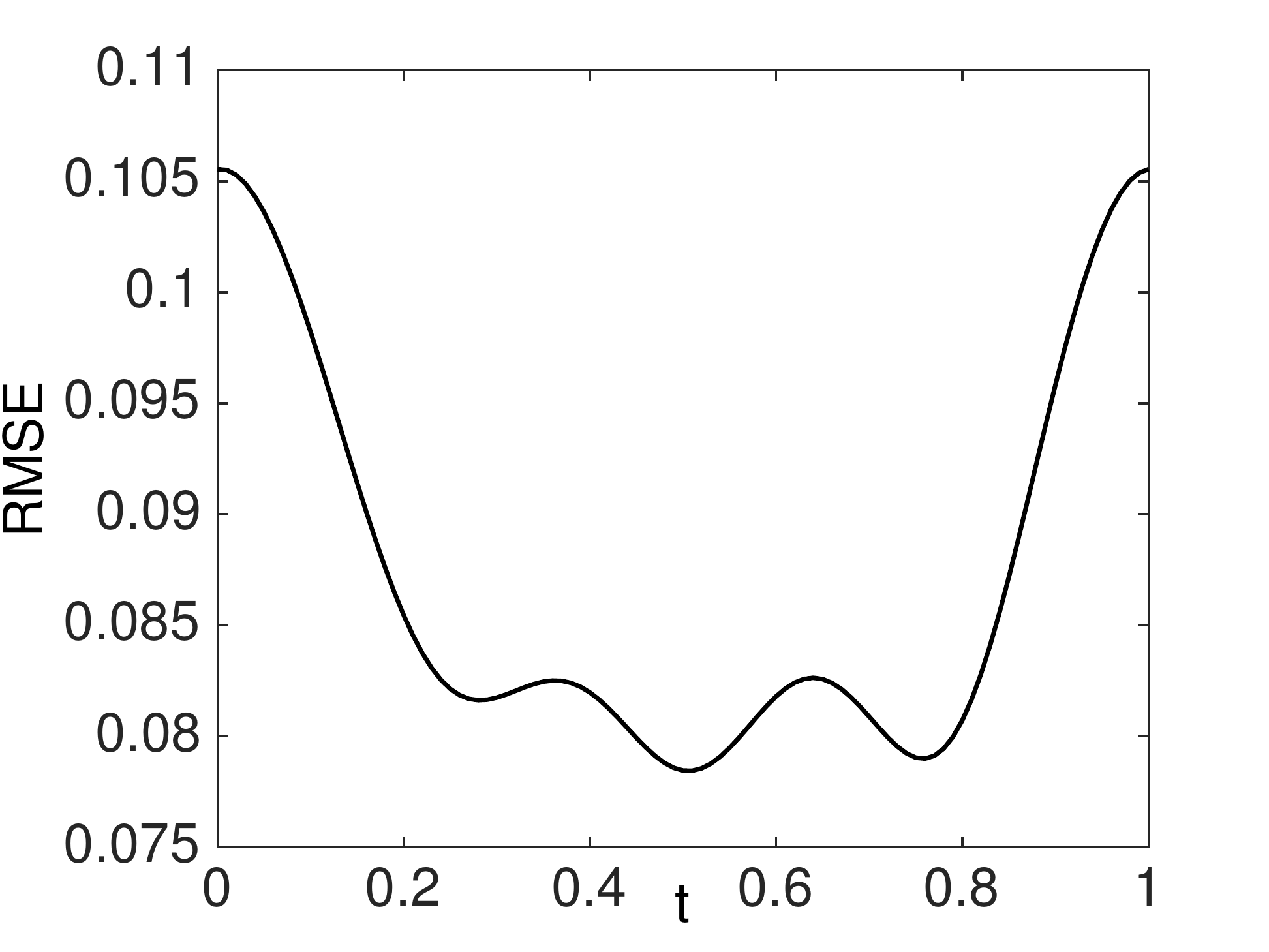}}
\end{minipage}
\caption{The pointwise mean, bias, standard deviation (STD) and root mean squared error (RMSE) of the estimated population slope function $\hat{\beta}(t)$ in 1,000 simulation replicates when $m_i=20$ and $\sigma_\epsilon=1.0$ in our simulation studies. The dashed line in the top left panel is the true  population slope function $\beta(t)=1+2\sin(2\pi t)+\cos(2\pi t)$.}
\label{fig:Case2}
\end{figure}

The functional linear mixed-effects model (\ref{Eqn:FLMR}) is estimated using the method introduced in Section 2. In case 1, we choose 35 cubic B-splines basis functions for $\beta(t)$ and $b_i(t)$; while in case 2, we choose 35 Fourier basis functions for $\beta(t)$ and $b_i(t)$. Figure \ref{fig:Case1} and \ref{fig:Case2} display the pointwise mean, bias, standard deviation and root mean squared error of the estimated population slope function $\beta(t)$ in 1,000 simulation replicates when $(m_i,n)=(20, 100)$ and $(\sigma_e,\sigma_\epsilon)=(0.5, 1.0)$. It shows that the pointwise mean of the estimated population slope function $\hat{\beta}(t)$ is very close to the true function $\beta(t)$ in both of cases.


The estimation results for all simulation setups are summarized in Table \ref{Table1}-\ref{Table2}. As expected, there is a substantial decrease in RMISE when more visits are observed per subject. When the functional covariate $X(t)$ is observed directly without measurement errors (i.e. $\sigma_e=0$), the mean and median of RMISE for the estimated population slope function $\hat{\beta}(t)$ and individual slope function $\hat{\beta}_i(t)$ is close, which indicates that the estimation is stable. In this case, when the number of replicated measurements for each individual increases from $m_i = 5$ to $m_i = 10$, the mean of RMISE of the estimated population slope function $\hat{\beta}(t)$ decreases 27\%, and the mean of RMISE of the estimated individual slope function $\hat{\beta}_i(t)$ decreases 8\%. When the functional covariate $X(t)$ is observed with measurement errors with the standard deviation $\sigma_e = 0.5$, the mean of RMISE of the estimated population slope function $\hat{\beta}(t)$ increases 36\%, and the mean of RMISE of the estimated individual slope function $\hat{\beta}_i(t)$ increases 25\%, in comparison with the case when the functional covariate $X(t)$ is observed directly without measurement errors.

\begin{table}[h]
\renewcommand{\arraystretch}{1.0}
\caption {The Bias, Standard deviation (STD) and RMSE of the intercept, and the means of RMISE of slope functions obtained by applying the REML-based EM algorithm on the simulated data with 1,000 simulation replicates. The true population slope function is $\beta(t)=1+2t^2+\exp(-3t)$.}
\label{Table1}
\begin{center}
\setlength{\tabcolsep}{0.9mm}
\begin{tabular}{ccccccccccccccc}
\hline
 \multirow{2}{*}{$\sigma_e$} & \multirow{2}{*}{} & \multirow{2}{*}{$\sigma_\epsilon$} &  \multirow{2}{*}{} & \multirow{2}{*}{$n$} &\multirow{2}{*}{} & \multirow{2}{*}{$m_i$} &  \multirow{2}{*}{}&\multicolumn{3}{c}{Intercept} &\multirow{2}{*}{} & \multirow{2}{*}{${\rm RMISE}\{\beta(t)\}$} & \multirow{2}{*}{} &  \multirow{2}{*}{${\rm RMISE}\{\beta_i(t)\}$}  \\
\cline{9-11}
& & & & & & & & Bias & STD & RMSE & & & & \\
\hline
0.0 &&  0.5& & 50& & 5 & & -0.0017&0.0986 &0.0986 & & 0.0047&&0.0210\\
    &&      & &   & & 10 & &0.0019&0.0926 &0.0926 & & 0.0032&&0.0187\\
    &&      & &   & & 20 & & 0.0005&0.0862 &0.0862 & & 0.0025&&0.0166\\
    &&      & &100& & 5 & & -0.0034&0.0638 &0.0639 & & 0.0027&&0.0180\\
    &&      & &   & & 10 & &-0.0012&0.0611 &0.0611 & & 0.0019&&0.0161\\
    &&      & &   & & 20 & & -0.0021&0.0609 &0.0610 & & 0.0014&&0.0145\\
\hline
0.0 && 1.0& &50 & & 5 & & -0.0010&0.1079 &0.1079 & & 0.0080&&0.0274\\
    &&      & &   & &10 & &0.0000&0.1023 &0.1023 & & 0.0060&&0.0237\\
    &&      & &   & &20 & & -0.0018&0.0923 &0.0923 & & 0.0043&&0.0204\\
    &&      & &100& &5 & & -0.0006&0.0760 &0.0760 & & 0.0054&&0.0230\\
    &&      & &   & &10 & &-0.0016&0.0663 &0.0663 & & 0.0036&&0.0198\\
    &&      & &   & &20 & & -0.0015&0.0630 &0.0630 & & 0.0024&&0.0174\\
\hline
0.5 && 0.5 & &50 & &5 & &  -0.0024&0.1001 &0.1002 & & 0.0049&&0.0212\\
    &&      & &   & &10 & & 0.0003&0.0952 &0.0952 & & 0.0035&&0.0189\\
    &&      & &   & &20 & & 0.0005&0.0890 &0.0890 & & 0.0033&&0.0177\\
    &&      & &100& &5 & &  -0.0023&0.0636 &0.0637 & & 0.0029&&0.0182\\
    &&      & &   & &10 & & 0.0011&0.0620 &0.0620 & & 0.0020&&0.0162\\
    &&      & &   & &20 & & -0.0016&0.0623 &0.0624 & & 0.0022&&0.0155\\
\hline
0.5 && 1.0& &50 & &5 & &  -0.0036&0.1130 &0.1131 & & 0.0082&&0.0277\\
    &&      & &   & &10 & & 0.0026&0.1039 &0.1039 & & 0.0057&&0.0235\\
    &&      & &   & &20 & & -0.0001&0.0935 &0.0935 & & 0.0043&&0.0204\\
    &&      & &100& &5 & &  -0.0023&0.0764 &0.0764 & & 0.0053&&0.0228\\
    &&      & &   & &10 & & 0.0009&0.0698 &0.0698 & & 0.0037&&0.0199\\
    &&      & &   & &20 & & -0.0011&0.0633 &0.0633 & & 0.0026&&0.0176\\
\hline
\end{tabular}
\end{center}
\end{table}


\begin{table}[h]
\renewcommand{\arraystretch}{1.0}
\caption {The Bias, Standard deviation (STD) and RMSE of the intercept, and the means of RMISE of slope functions obtained by applying the REML-based EM algorithm on the simulated data with 1,000 simulation replicates. The true population slope function is $\beta(t)=1+2\sin(2\pi t)+\cos(2\pi t)$.}
\label{Table2}
\begin{center}
\setlength{\tabcolsep}{0.9mm}
\begin{tabular}{ccccccccccccccc}
\hline
\multirow{2}{*}{$\sigma_e$} & \multirow{2}{*}{} & \multirow{2}{*}{$\sigma_\epsilon$} &  \multirow{2}{*}{} & \multirow{2}{*}{$n$} &\multirow{2}{*}{} & \multirow{2}{*}{$m_i$} &  \multirow{2}{*}{}&\multicolumn{3}{c}{Intercept} &\multirow{2}{*}{} & \multirow{2}{*}{${\rm RMISE}\{\beta(t)\}$} & \multirow{2}{*}{} &  \multirow{2}{*}{${\rm RMISE}\{\beta_i(t)\}$}  \\
\cline{9-11}
& & & & & & & & Bias & STD & RMSE & & && \\
\hline
0.0 && 0.5& & 50 & &5 & & 0.0039&0.0980 &0.0981 & & 0.0049&&0.0415\\
    &&      & &    & &10 & &0.0032&0.0912 &0.0912 & & 0.0038&&0.0409\\
    &&      & &    & &20 & & -0.0012&0.0890 &0.0890 & & 0.0033&&0.0402\\
    &&       & &100 & &5 & & -0.0014&0.0643 &0.0644 & & 0.0026&&0.0405\\
    &&       & &    & &10 & &-0.0014&0.0630 &0.0631 & & 0.0023&&0.0396\\
    &&       & &    & &20 & & -0.0008&0.0629 &0.0629 & & 0.0019&&0.0395\\
\hline
0.0  && 1.0& & 50 & &5 & & 0.0059&0.1106 &0.1107 & & 0.0082&&0.0447\\
     &&      & &    & &10 & &0.0050&0.1005 &0.1006 & & 0.0053&&0.0423\\
     &&      & &    & &20 & & 0.0006&0.0948 &0.0948 & & 0.0038&&0.0407\\
     &&      & &100 & &5 & & -0.0014&0.0747 &0.0747 & & 0.0039&&0.0418\\
     &&      & &    & &10 & &-0.0006&0.0687 &0.0687 & & 0.0028&&0.0401\\
     &&      & &    & &20 & & -0.0003&0.0653 &0.0653 & & 0.0022&&0.0397\\
\hline
0.5  && 0.5& &50  & &5 & &  0.0021&0.0981 &0.0981 & & 0.0052&&0.0418\\
     &&      & &    & &10 & & 0.0042&0.0929 &0.0930 & & 0.0039&&0.0409\\
     &&      & &    & &20 & & -0.0012&0.0895 &0.0895 & & 0.0033&&0.0402\\
     &&      & &100 & &5 & &  -0.0030&0.0643 &0.0644 & & 0.0027&&0.0406\\
     &&      & &    & &10 & & -0.0016&0.0620 &0.0620 & & 0.0023&&0.0396\\
     &&      & &    & &20 & & -0.0013&0.0625 &0.0625 & & 0.0019&&0.0395\\
\hline
0.5 && 1.0& &50  & &5 & &  0.0026&0.1142 &0.1142 & & 0.0082&&0.0447\\
    &&       & &    & &10 & & 0.0026&0.1008 &0.1009 & & 0.0052&&0.0422\\
    &&       & &    & &20 & & -0.0020&0.0927 &0.0927 & & 0.0039&&0.0408\\
    &&       & &100 & &5 & &  -0.0034&0.0754 &0.0755 & & 0.0040&&0.0419\\
    &&       & &    & &10 & & -0.0015&0.0678 &0.0678 & & 0.0029&&0.0402\\
    &&       & &    & &20 & & 0.0004&0.0652 &0.0652 & & 0.0022&&0.0397\\
\hline
\end{tabular}
\end{center}
\end{table}

\vskip 0.5cm

\section{Applications}

In this section, we perform the afore-proposed FLMMs via the EM algorithm to analyze two applications.

\subsection{Ozone Pollution Analysis}
The first study is to re-visit the air pollution study introduced in Section 1. The functional linear mixed-effects model (2) is used to study the effect of the 24-hour nitrogen dioxide $NO_2$(t) on the daily maximum ozone concentration. We fit the following mixed-effect model
\begin{eqnarray*}
Y_{ij} &=& \alpha_0+a_i+\int_0^{23} [\beta(t)+b_i(t)] X_{ij}(t){\rm d}t+\epsilon_{ij},\\
& & a_i \sim N(0, \sigma^2_a),~ \epsilon_{ij} \sim N(0, \sigma^2_\epsilon),~i=1,...,n, ~j=1,..., m_i\;,
\end{eqnarray*}
where $Y_{ij}$ is the daily maximum ozone within 24 hours for the $j$-th day in the $i$-th city, $X_{ij}(t)$ is the 24-hour nitrogen dioxide $NO_2$(t) observations. The data are collected for $n=62$ cities from April 13 to September 30, 1996.

For computational facilities, we have considered the penalized spline estimator and expand the functional coefficients in cubic
B-splines basis functions with $K = 26$ equispaced interior knots for the population slope
function $\beta(t)$ and the random slope function $b_i(t)$.

The fixed effects $\{\alpha_0, \beta(t)\}$ and random effects $\{a_i, b_i(t)\}$ are estimated by minimizing
\begin{eqnarray*}
\label{eqn:LogLIK}
&& H(\btheta,\bxi) \\
&=& \sum\limits_{i=1}^n \sum\limits_{j=1}^{m_i}  \frac{1}{2\sigma^2_\epsilon}\bigg(Y_{ij} - \alpha_0 - a_i - \int_0^{23} [\beta(t)+b_i(t)] X_{ij}(t) {\rm d}t \bigg)^2+\frac{1}{2}\sum\limits_{i=1}^n\bb'_i\bD^{-1}\bb_i \nonumber\\
&& + \bigg[\frac{\lambda_\beta}{2} \int_0^{23}  \bigg\{\frac{{\rm d}^2\beta(t)}{{\rm d}t^2}\bigg\}^2{\rm d}t + \frac{\lambda_b}{2} \sum_{i=1}^n \int_0^{23}  \bigg\{ \frac{{\rm d}^2b_i(t)}{{\rm d}t^2} \bigg\}^2{\rm d}t \bigg] + \frac{1}{2\sigma^2_a}\sum_{i=1}^n a^2_i\;.
\end{eqnarray*}
The smoothing parameters are chosen as $\lambda_\beta = 10^{2.0}$ and $\lambda_b = 10^{0.5} $ by GCV criterion. We implement the REML-based EM algorithm proposed in Section 2.4. The estimate for the intercept $\alpha_0$ is $\hat{\alpha}_0=3.8262$ with the estimated standard error 0.0094, and the 95\% confidence interval of $\alpha_0$ is $[3.8078,~3.8446]$.

\begin{figure}
\begin{center}
\includegraphics[width=12.0cm, height=10.0cm]{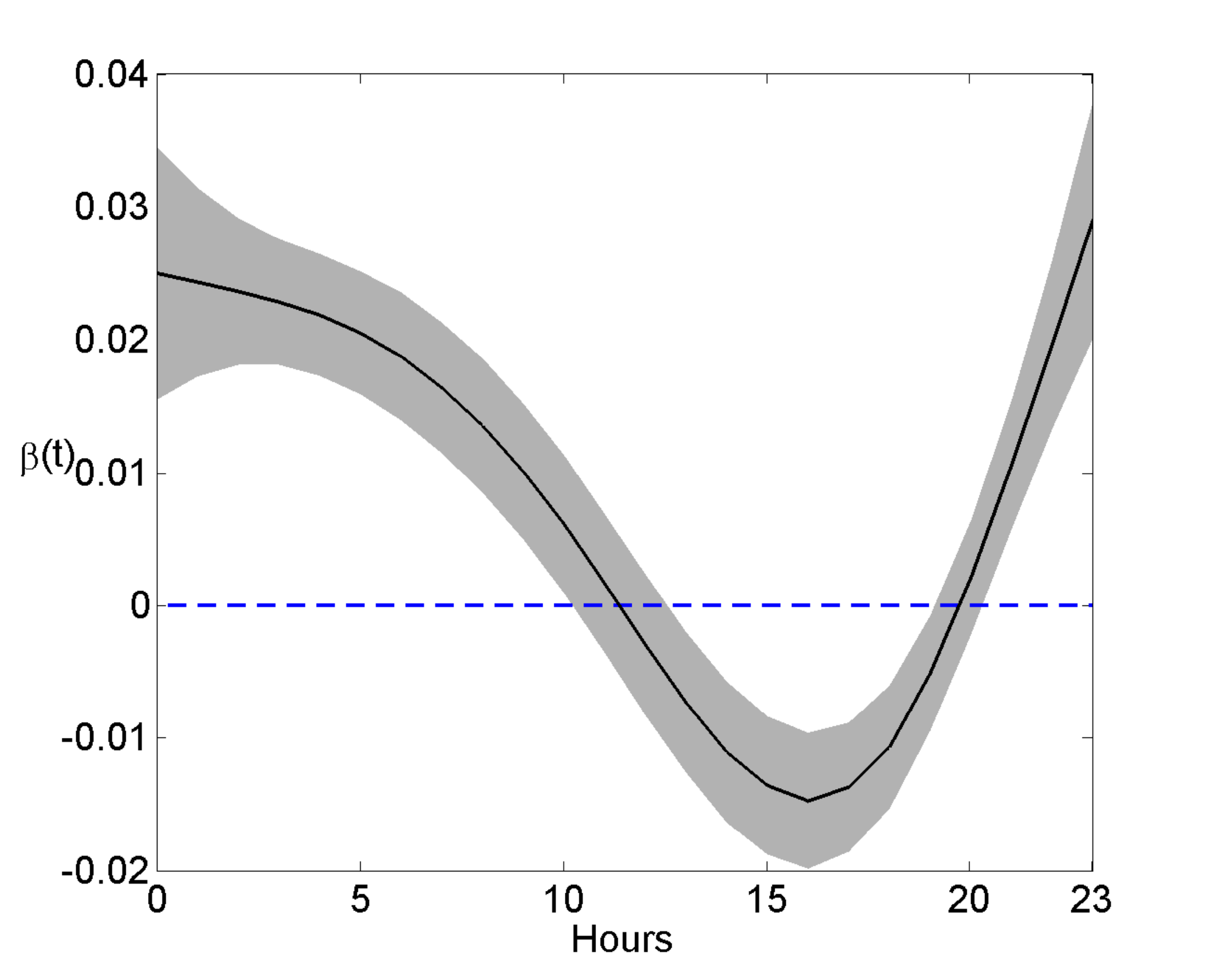}
\caption {The estimated population slope function $\hat{\beta}(t)$ for predicting the log maximum of ozone concentration from the hourly $NO_2(t)$. The shaded area indicates the pointwise 95\% confidence interval for $\hat{\beta}(t)$.}
\label{Fig:O3AveSlope}
\end{center}
\end{figure}

Figure \ref{Fig:O3AveSlope} displays the estimate population slope function $\hat{\beta}(t)$ and its approximate 95\% pointwise confidence interval. We can see a positive  correlation between the maximum ozone concentration and the nitrogen dioxide before 11 am and after 8 pm but negative correlation between 11 am and 8 pm. Due to the stopping of the sun lighting in the night, a lot of nitrogen dioxide is accumulated; with the sunrise at about 6 am, more and more nitrogen dioxide is reacted with the sun light to generate ozone, so more ozone is generated with the decreasing of nitrogen dioxide. This process will last until the sunset at about 7-8 pm, then the nitrogen dioxide is accumulated again.

\begin{figure}
\begin{minipage}{0.48\linewidth}
\centerline{\includegraphics[width=6.0cm]{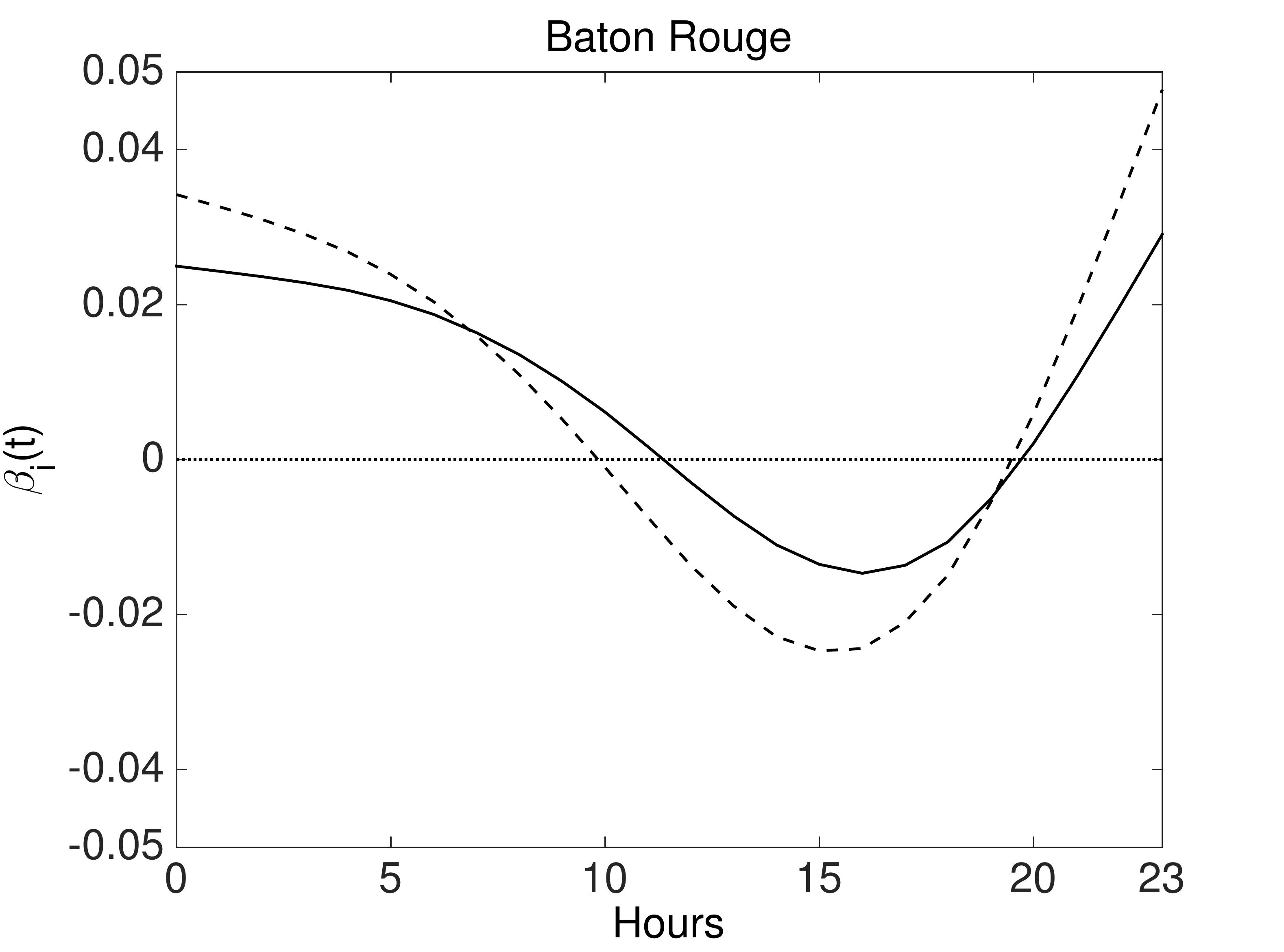}}
\end{minipage}
\hfill
\begin{minipage}{.48\linewidth}
\centerline{\includegraphics[width=6.0cm]{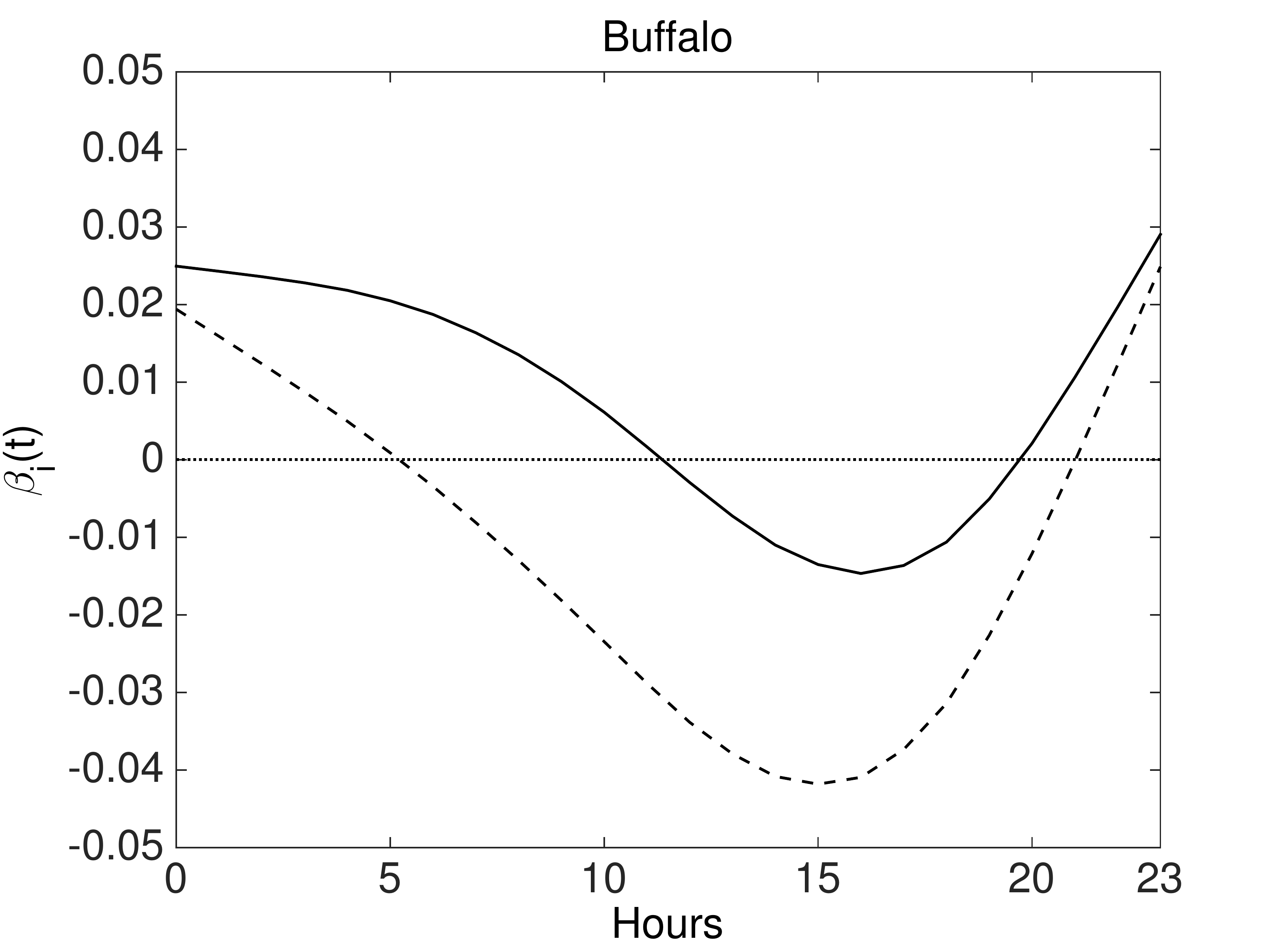}}
\end{minipage}
\vfill
\begin{minipage}{0.48\linewidth}
\centerline{\includegraphics[width=6.0cm]{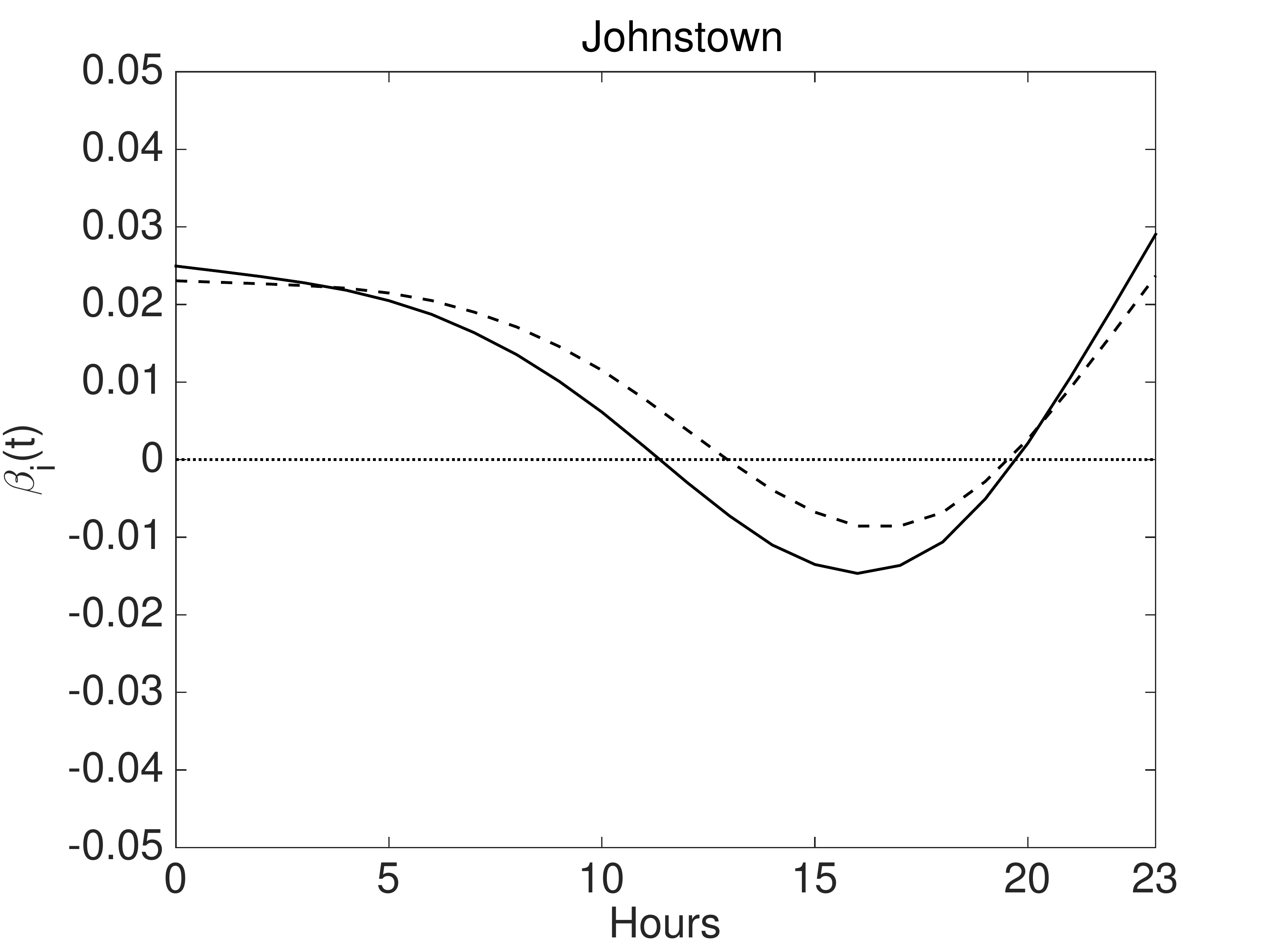}}
\end{minipage}
\hfill
\begin{minipage}{0.48\linewidth}
\centerline{\includegraphics[width=6.0cm]{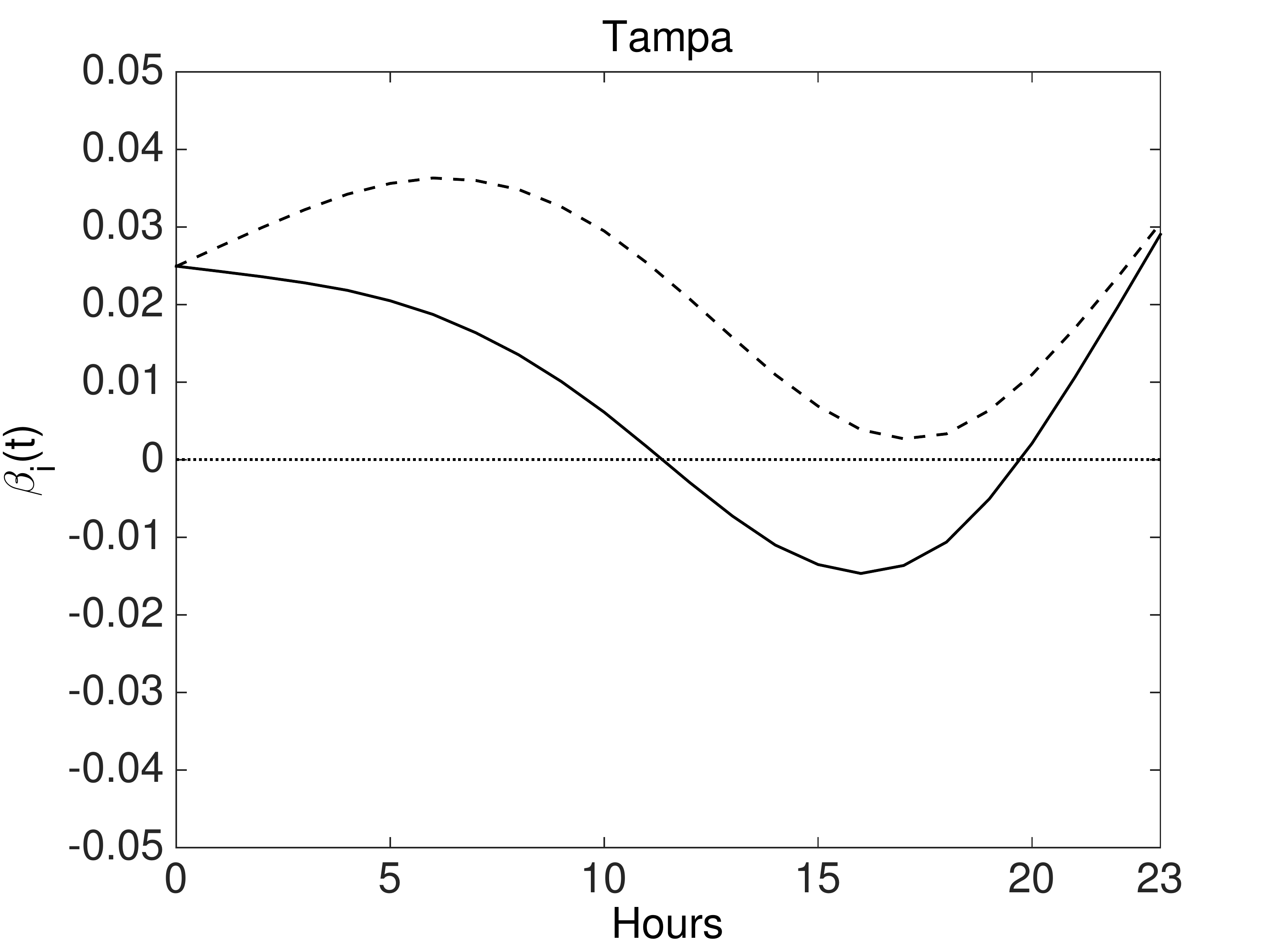}}
\end{minipage}
\caption {The estimated individual slope function $\hat{\beta}_i(t) = \hat{\beta}(t) + \hat{b}_i(t)$ for predicting the log maximum ozone concentration from the hourly $NO_2(t)$ for four cities:  Baton Rouge, Buffalo, Johnstown, and Tampa. The solid line is the estimated population slope function $\hat{\beta}(t)$, and the dashed line is the individual slope function $\hat{\beta}_i(t)$.}
\label{Fig:O3SubjSlope}
\end{figure}

Figure \ref{Fig:O3SubjSlope} displays the estimated individual slope function $\hat{\beta}_i(t) = \hat{\beta}(t) + \hat{b}_i(t)$ for four cities: Baton Rouge, Buffalo, Johnstown, and Tampa. The individual slope function $\hat{\beta}_i(t)$ for Buffalo is lower than the population slope function in the whole day, which indicates that the hourly nitrogen dioxide has a lower effect on maximum ozone concentration in the whole day. On the other hand, the individual slope function $\hat{\beta}_i(t)$ for Tampa is higher than the population slope function in the whole day. This interesting phenomenon cannot be found from the regular functional linear regression model.

%
%

\subsection{Weather Data Analysis}
In this study, we are exploring the effect of the daily temperature in each year on the annual precipitation. We use the dataset consisting of the annual precipitation and the corresponding daily temperature measurements for 38 Canadian weather stations in 1961-1991. There are a lot of missing data in the year 1979, thus we delete the data in the year 1979. The functional linear mixed-effects model of our interest is
$$
Y_{ij}=\alpha_0+a_i+\int_0^{365} [\beta(t)+b_i(t)] X_{ij}(t){\rm d}t+\epsilon_{ij},~~~~ a_i \sim N(0, \sigma^2_a),~~ \epsilon_{ij} \sim N(0, \sigma^2_\epsilon)\;,
$$
where $Y_{ij}$ is the logarithm of annual precipitation at the $i$-th weather station in the $j$-th year, and $X_{ij}(t)$ is the daily temperature profile for $i=1,\ldots,38$, $j=1,\ldots,30$.

Due to the periodicity of weather data, we choose 35 Fourier basis functions to represent the population slope function $\beta(t)$ and the individual slope function $b_i(t)$. We follow the suggestion of Ramsay and Silverman (2005) to use the harmonic acceleration operator to define the roughness penalty for the population slope function $\beta(t)$. The harmonic acceleration operator is defined as $L\beta(t) ={{\rm d}^3\beta(t)}/{{\rm d}t^3} + \omega^2 {{\rm d}\beta(t)}/{{\rm d}t}$, where $\omega=\frac{2\pi}{365}$ is the period of the nonparametric function. Therefore, the zero roughness implies that $\beta(t)$ is of the form $\beta(t) = a_1+a_2 \sin(\omega t) + a_3 \cos(\omega t)$. The harmonic acceleration operator is also used to define the roughness penalty of the individual slope function $b_i(t)$.

The fixed effects $\{\alpha_0, \beta(t)\}$ and random effects $\{a_i, b_i(t)\}$ are estimated by minimizing
\begin{eqnarray*}
&& H(\btheta,\bxi) \\
&=& \sum\limits_{i=1}^n \sum\limits_{j=1}^{m_i}  \frac{1}{2\sigma^2_\epsilon}\bigg(Y_{ij} - \alpha_0 - a_i - \int_0^{365} [\beta(t)+b_i(t)] X_{ij}(t) {\rm d}t \bigg)^2+\frac{1}{2}\sum\limits_{i=1}^n\bb'_i\bD^{-1}\bb_i \\
&& + \bigg[\frac{\lambda_\beta}{2} \int_0^{365}  \bigg\{L\beta(t)\bigg\}^2{\rm d}t + \frac{\lambda_b}{2} \sum_{i=1}^n \int_0^{365}  \bigg\{L b_i(t) \bigg\}^2{\rm d}t \bigg] + \frac{1}{2\sigma^2_a}\sum_{i=1}^n a^2_i\;.
\end{eqnarray*}
The smoothing parameters are chosen as $\lambda_\beta = 10^{13.75}$ and $\lambda_b = 10^{12.25} $ by GCV criterion. We implement the REML-based EM algorithm proposed in Section 2.2. The estimate for the intercept $\alpha_0$ is $\hat{\alpha}_0=2.994$ with the estimated standard error $0.055$, and the 95\% confidence interval of $\alpha_0$ is $[2.886,~3.101]$.

\begin{figure}
\begin{center}
\includegraphics[width=12.0cm, height=10.0cm]{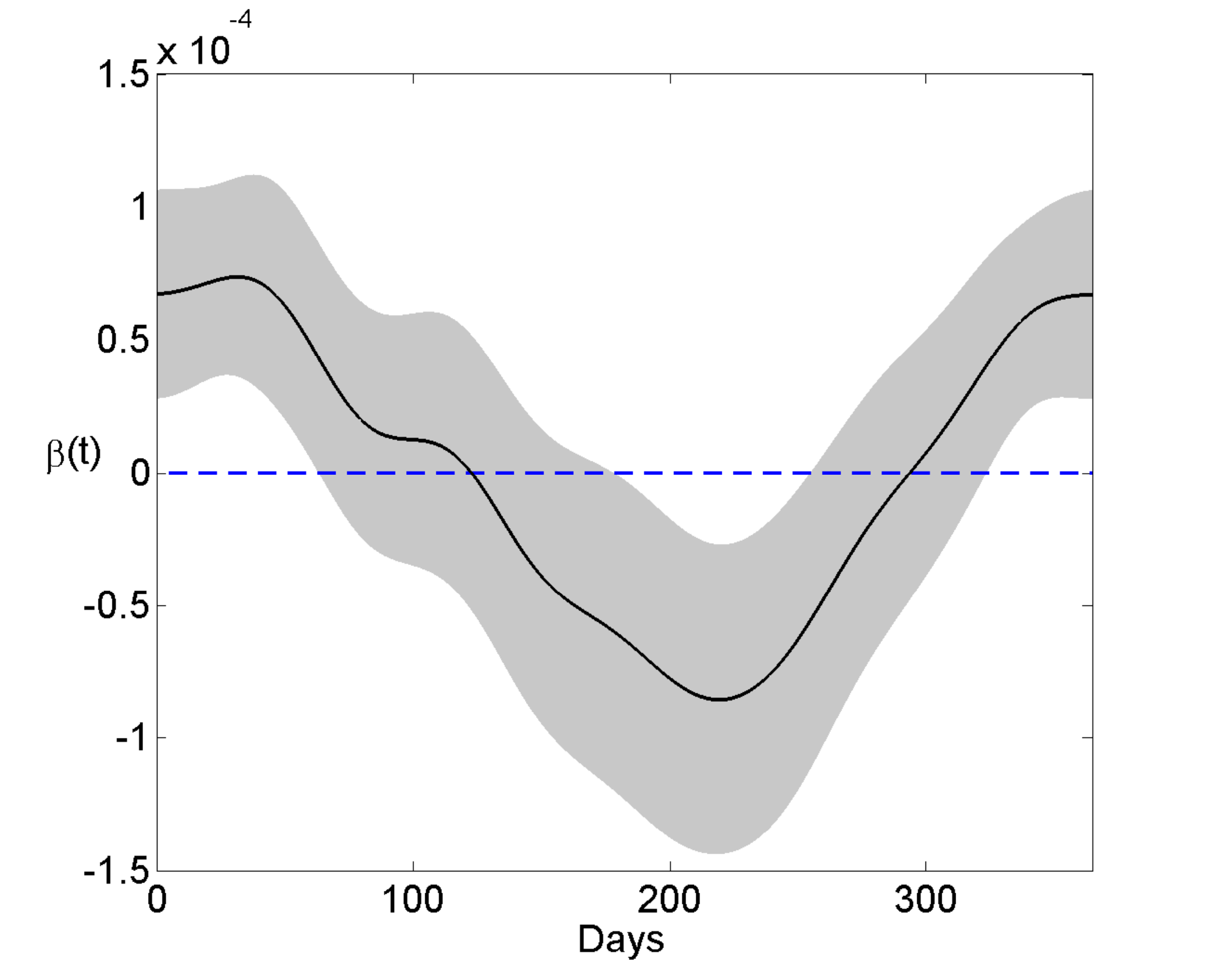}
\caption {The estimated population slope function $\hat{\beta}(t)$ for predicting the log total annual precipitation from the daily temperature. The shaded area indicates the pointwise 95\% confidence interval for $\hat{\beta}(t)$.}
\label{Fig:AveSlope}
\end{center}
\end{figure}

Figure \ref{Fig:AveSlope} displays the estimated population slope function $\beta(t)$ and the 95\% pointwise confidence interval. It indicates that the temperature in the winter has a significant and positive effect on the annual precipitation. The temperature in the summer has a negative effect on the annual precipitation, but this effect is only marginally significant.

\begin{figure}[h]
\begin{minipage}{0.48\linewidth}
\centerline{\includegraphics[width=6.0cm]{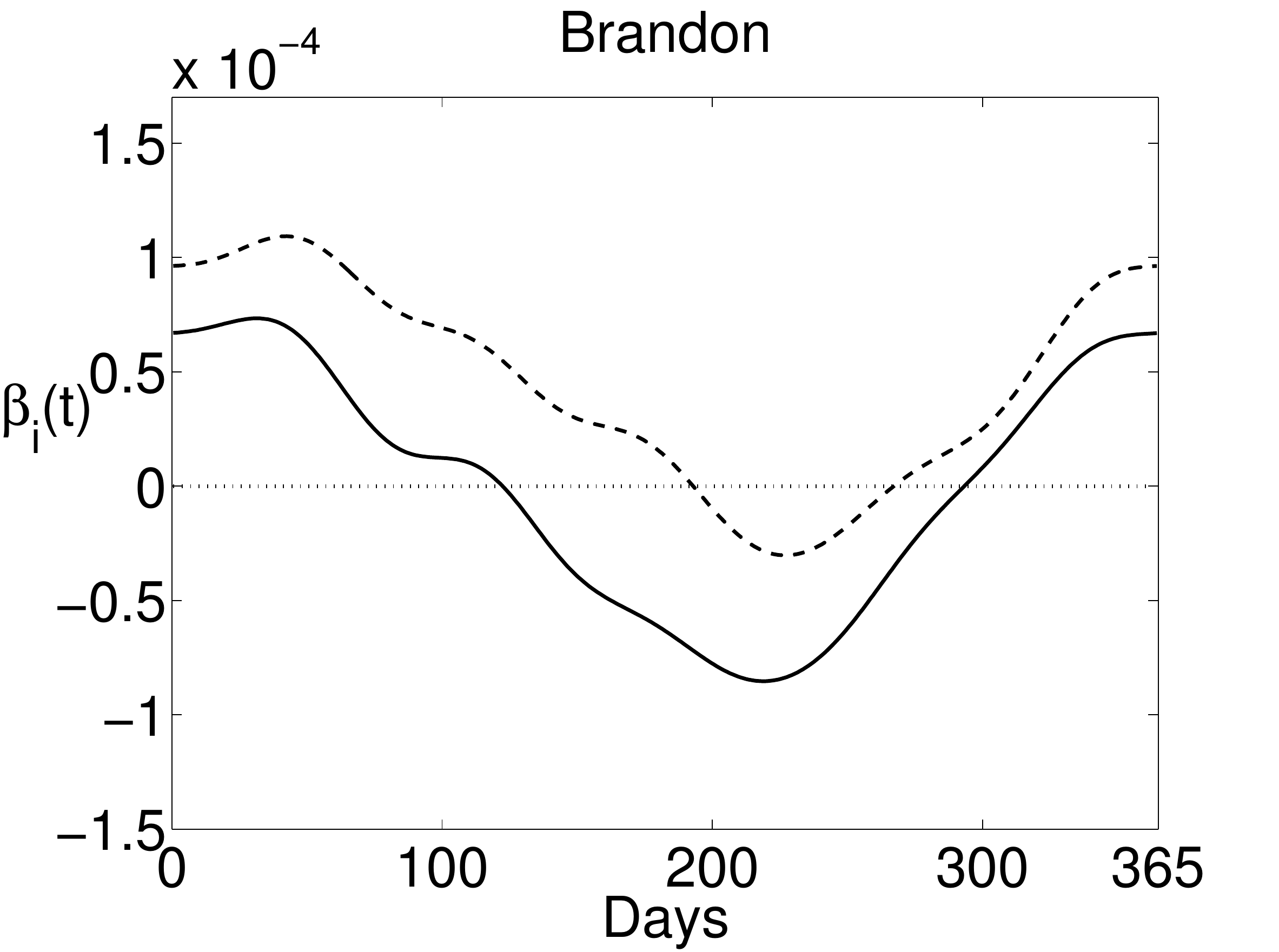}}
\end{minipage}
\hfill
\begin{minipage}{.48\linewidth}
\centerline{\includegraphics[width=6.0cm]{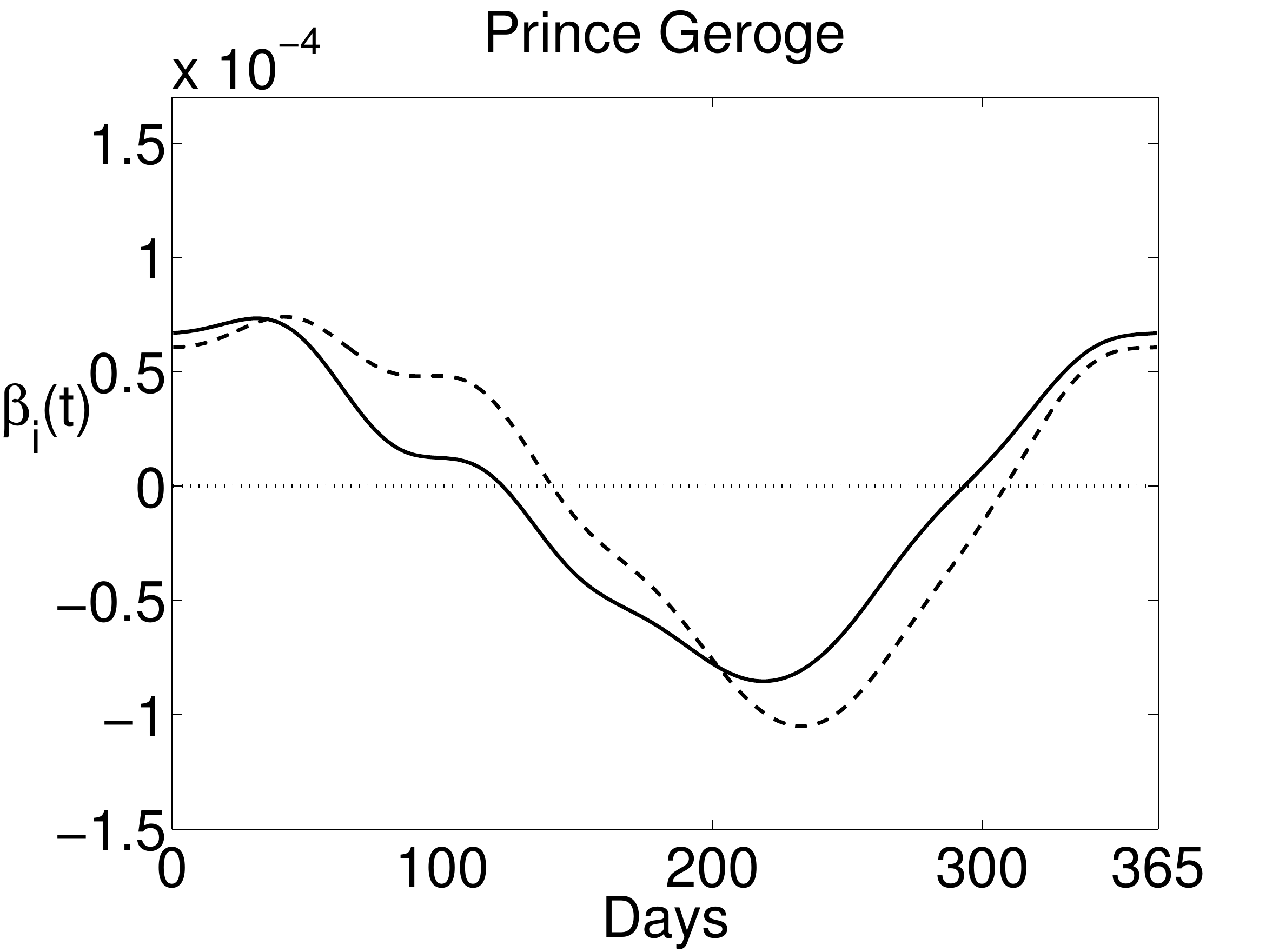}}
\end{minipage}
\vfill
\begin{minipage}{0.48\linewidth}
\centerline{\includegraphics[width=6.0cm]{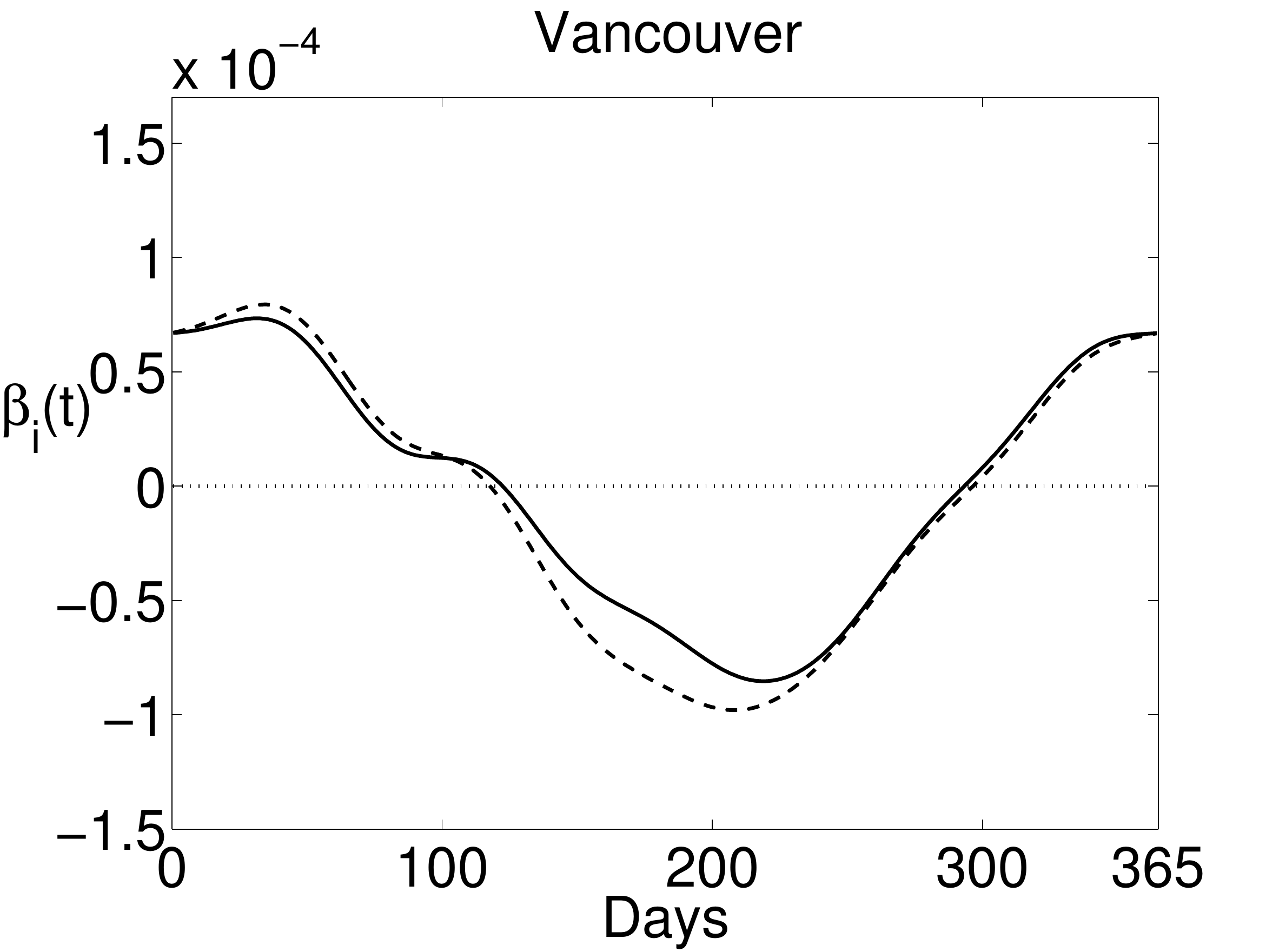}}
\end{minipage}
\hfill
\begin{minipage}{0.48\linewidth}
\centerline{\includegraphics[width=6.0cm]{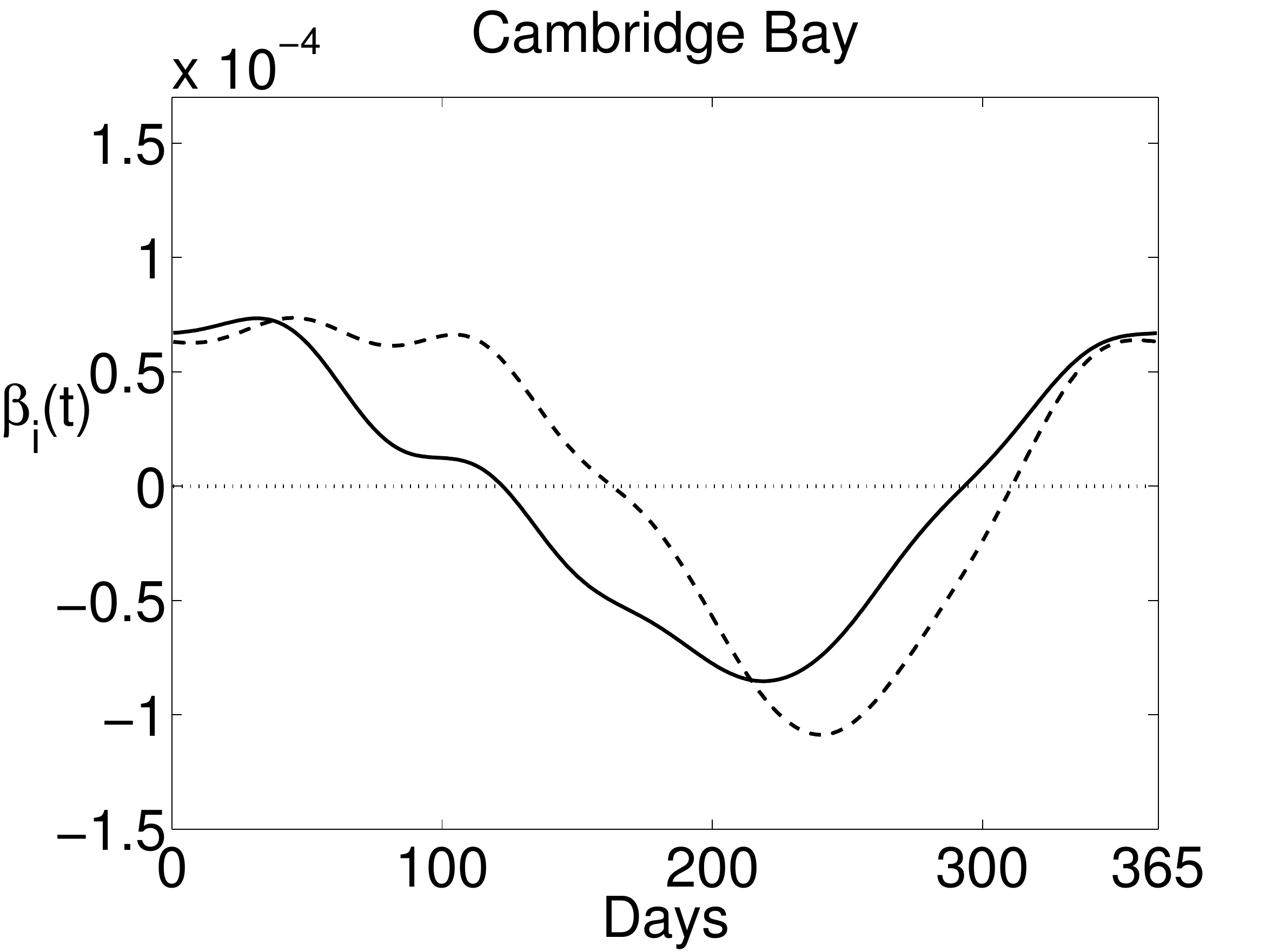}}
\end{minipage}
\caption {The estimated individual slope function $\hat{\beta}_i(t) = \hat{\beta}(t) + \hat{b}_i(t)$ for predicting the log total annual precipitation from the daily temperature for four cities: Brandon, Prince George, Vancouver, Cambridge Bay. The solid line is the estimated population slope function $\hat{\beta}(t)$, and the dashed line is the individual slope function $\hat{\beta}_i(t)$. }
\label{Fig:IndiSlope}
\end{figure}

We also plot the individual slope function $\hat{\beta}_i(t) = \hat{\beta}(t) + \hat{b}_i(t)$ for four stations in Figure \ref{Fig:IndiSlope}. There are some obvious individual variations from the population slope function for each station. For example, the Brandon city is located in western Manitoba province, on the banks of the Assiniboine River. Figure \ref{Fig:IndiSlope} shows that the individual slope function of Brandon is higher than the population slope function in the whole year, because Brandon has a lower latitude and a large amount of precipitation in most of whole year. This phenomenon cannot be inferred from the regular functional linear regression model.

%
%

\section{Conclusions}
The functional linear regression model (\ref{Eqn:FLM}) is a popular tool to analyze the relationship between a scalar response variable and a functional covariate. But when there are repeated measurements on multiple subjects, the same slope function seems to be a too strict assumption. In this article, we propose a functional linear mixed-effect model (\ref{Eqn:FLMR}). This model is more flexible than the regular functional linear regression model in the sense that each subject has their individual slope function while all subjects share a population slope function.

The population and random slope functions are estimated by the penalized spline smoothing method, in which the roughness of the slope functions are controlled by a penalty function. The variance parameters for the random slope function and the data noise are estimated by a REML-based EM algorithm. Our simulation studies show that the estimation method can provide accurate estimates for the functional linear mixed-effect model.

The functional linear mixed-effect model is demonstrated using two real applications. The first application uses the functional linear mixed-effects model (\ref{Eqn:FLMR}) to study the effect of the 24-hour nitrogen dioxide on the daily maximum ozone concentration. Some interesting results are found. For example, the hourly nitrogen dioxide has a consistently higher effect on the daily maximum ozone concentration in the whole day in some cities such as Tampa. These insights cannot be gained from the regular functional linear regression models. The similar phenomenon is also found in our second application to investigate the effect of the daily temperature on the annual precipitation.

\section*{References}

\bibliographystyle{model2-names}
\bibliography{thesisJiguo}

\begin{thebibliography}{19}
\expandafter\ifx\csname natexlab\endcsname\relax\def\natexlab#1{#1}\fi
\providecommand{\url}[1]{\texttt{#1}}
\providecommand{\href}[2]{#2}
\providecommand{\path}[1]{#1}
\providecommand{\DOIprefix}{doi:}
\providecommand{\ArXivprefix}{arXiv:}
\providecommand{\URLprefix}{URL: }
\providecommand{\Pubmedprefix}{pmid:}
\providecommand{\doi}[1]{\href{http://dx.doi.org/#1}{\path{#1}}}
\providecommand{\Pubmed}[1]{\href{pmid:#1}{\path{#1}}}
\providecommand{\bibinfo}[2]{#2}
\ifx\xfnm\relax \def\xfnm[#1]{\unskip,\space#1}\fi
\bibitem[{Ash and Gardner(1975)}]{Ash75}
\bibinfo{author}{Ash, R.B.}, \bibinfo{author}{Gardner, M.F.},
  \bibinfo{year}{1975}.
\newblock \bibinfo{title}{Topics in Stochastic Processes. Probability and
  Mathematical Statistics 27.}
\newblock \bibinfo{publisher}{Academic Press}, \bibinfo{address}{New York}.
\bibitem[{Cai and Hall(2006)}]{CaiHall06}
\bibinfo{author}{Cai, T.T.}, \bibinfo{author}{Hall, P.}, \bibinfo{year}{2006}.
\newblock \bibinfo{title}{Prediction in functional linear regression}.
\newblock \bibinfo{journal}{The Annals of Statistics} \bibinfo{volume}{34},
  \bibinfo{pages}{2159--2179}.
\bibitem[{Cardot et~al.(2007)Cardot, Crambes and Sarda}]{Cardot07}
\bibinfo{author}{Cardot, H.}, \bibinfo{author}{Crambes, C.},
  \bibinfo{author}{Sarda, P.}, \bibinfo{year}{2007}.
\newblock \bibinfo{title}{Ozone pollution forecasting using conditional mean
  and conditional quantiles with functional covariates}, in:
  \bibinfo{editor}{Hardle, W.}, \bibinfo{editor}{Mori, Y.},
  \bibinfo{editor}{Vieu, P.} (Eds.), \bibinfo{booktitle}{Statistical Methods
  for Biostatisticsand Related Fields}. \bibinfo{publisher}{Springer},
  \bibinfo{address}{Berlin}, pp. \bibinfo{pages}{221--243}.
\bibitem[{Chiou et~al.(2003)Chiou, M\"{u}ller and Wang}]{ChiouMullerWang03}
\bibinfo{author}{Chiou, J.M.}, \bibinfo{author}{M\"{u}ller, H.G.},
  \bibinfo{author}{Wang, J.L.}, \bibinfo{year}{2003}.
\newblock \bibinfo{title}{Functional quasi-likelihood regression models with
  smooth random effects}.
\newblock \bibinfo{journal}{J. R. Stat. Soc. Ser. B.} \bibinfo{volume}{65},
  \bibinfo{pages}{405--423}.
\bibitem[{Crambes et~al.(2009)Crambes, Kneip and Sarda}]{Crambes09}
\bibinfo{author}{Crambes, C.}, \bibinfo{author}{Kneip, A.},
  \bibinfo{author}{Sarda, P.}, \bibinfo{year}{2009}.
\newblock \bibinfo{title}{Smoothing splines estimators for functional linear
  regression}.
\newblock \bibinfo{journal}{The Annals of Statistics} \bibinfo{volume}{37},
  \bibinfo{pages}{35--72}.
\bibitem[{Ferraty and Vieu(2006)}]{FerratyVieu06}
\bibinfo{author}{Ferraty, F.}, \bibinfo{author}{Vieu, P.},
  \bibinfo{year}{2006}.
\newblock \bibinfo{title}{Nonparametric Functional Data Analysis: Methods,
  Theory, Applications and Implementations}.
\newblock \bibinfo{publisher}{Springer-Verlag}, \bibinfo{address}{London}.
\bibitem[{Goldsmith et~al.(2012)Goldsmith, Crainiceanu, Caffo and
  Reich}]{Goldsmith12}
\bibinfo{author}{Goldsmith, J.}, \bibinfo{author}{Crainiceanu, C.M.},
  \bibinfo{author}{Caffo, B.}, \bibinfo{author}{Reich, D.},
  \bibinfo{year}{2012}.
\newblock \bibinfo{title}{Longitudinal penalized functional regression for
  cognitive outcomes on neuronal tract measurements}.
\newblock \bibinfo{journal}{Journal of the Royal Statistical Society, Series C}
  \bibinfo{volume}{61}, \bibinfo{pages}{453--469}.
\bibitem[{Goldsmith et~al.(2011)Goldsmith, Wand and Crainiceanu}]{Goldsmith11}
\bibinfo{author}{Goldsmith, J.}, \bibinfo{author}{Wand, M.P.},
  \bibinfo{author}{Crainiceanu, C.}, \bibinfo{year}{2011}.
\newblock \bibinfo{title}{Functional regression via variational bayes}.
\newblock \bibinfo{journal}{Electronic Journal of Statistics}
  \bibinfo{volume}{5}, \bibinfo{pages}{572--602}.
\bibitem[{Morris(2015)}]{Morris15}
\bibinfo{author}{Morris, J.S.}, \bibinfo{year}{2015}.
\newblock \bibinfo{title}{Functional regression}.
\newblock \bibinfo{journal}{Statistics and Its Application}
  \bibinfo{volume}{2}, \bibinfo{pages}{321--359}.
\bibitem[{Peng and Welty(2004)}]{Peng2004}
\bibinfo{author}{Peng, R.}, \bibinfo{author}{Welty, L.}, \bibinfo{year}{2004}.
\newblock \bibinfo{title}{The nmmapsdata package}.
\newblock \bibinfo{journal}{R News} \bibinfo{volume}{4},
  \bibinfo{pages}{10--14}.
\bibitem[{Ramsay and Dalzell(1991)}]{RamsayDalzell91}
\bibinfo{author}{Ramsay, J.O.}, \bibinfo{author}{Dalzell, C.J.},
  \bibinfo{year}{1991}.
\newblock \bibinfo{title}{Some tools for functional data analysis}.
\newblock \bibinfo{journal}{Journal of the Royal Statistical Society, Series B}
  \bibinfo{volume}{53}, \bibinfo{pages}{539--572}.
\bibitem[{Ramsay and Silverman(2002)}]{RamsaySilverman02}
\bibinfo{author}{Ramsay, J.O.}, \bibinfo{author}{Silverman, B.W.},
  \bibinfo{year}{2002}.
\newblock \bibinfo{title}{Applied Functional Data Analysis}.
\newblock \bibinfo{publisher}{Springer}, \bibinfo{address}{New York}.
\bibitem[{Ramsay and Silverman(2005)}]{RamsaySilverman05}
\bibinfo{author}{Ramsay, J.O.}, \bibinfo{author}{Silverman, B.W.},
  \bibinfo{year}{2005}.
\newblock \bibinfo{title}{Functional Data Analysis}.
\newblock \bibinfo{edition}{Second} ed., \bibinfo{publisher}{Springer},
  \bibinfo{address}{New York}.
\bibitem[{Rice and Silverman(1991)}]{Rice1991}
\bibinfo{author}{Rice, J.A.}, \bibinfo{author}{Silverman, B.W.},
  \bibinfo{year}{1991}.
\newblock \bibinfo{title}{Estimating the mean and covariance structure
  nonparametrically when the data are curves}.
\newblock \bibinfo{journal}{Journal of the Royal Statistical Society, Series B}
  \bibinfo{volume}{53}, \bibinfo{pages}{233--243}.
\bibitem[{Staniswalis and Lee(1998)}]{StaniswalisLee98}
\bibinfo{author}{Staniswalis, J.G.}, \bibinfo{author}{Lee, J.J.},
  \bibinfo{year}{1998}.
\newblock \bibinfo{title}{Nonparametric regression analysis of longitudinal
  data}.
\newblock \bibinfo{journal}{Journal of the American Statistical Association}
  \bibinfo{volume}{93}, \bibinfo{pages}{1403--1418}.
\bibitem[{Wu and Zhang(2006)}]{WuZhang2006}
\bibinfo{author}{Wu, H.L.}, \bibinfo{author}{Zhang, J.T.},
  \bibinfo{year}{2006}.
\newblock \bibinfo{title}{Nonparametric Regression Methods for Longitudinal
  Data Analysis: Mixed-Effects Modeling Approaches}.
\newblock \bibinfo{publisher}{Wiley}, \bibinfo{address}{New York}.
\bibitem[{Wu et~al.(2010)Wu, Fan and M{\"u}ller}]{WuFan2010}
\bibinfo{author}{Wu, Y.C.}, \bibinfo{author}{Fan, J.Q.},
  \bibinfo{author}{M{\"u}ller, H.G.}, \bibinfo{year}{2010}.
\newblock \bibinfo{title}{Varying-coefficient functional linear regression}.
\newblock \bibinfo{journal}{Bernoulli} \bibinfo{volume}{16},
  \bibinfo{pages}{730--758}.
\bibitem[{Yao et~al.(2005)Yao, M{\"u}ller and Wang}]{Yao2005}
\bibinfo{author}{Yao, F.}, \bibinfo{author}{M{\"u}ller, H.},
  \bibinfo{author}{Wang, J.}, \bibinfo{year}{2005}.
\newblock \bibinfo{title}{Fucntional linear regression analysis for
  longitudinal data}.
\newblock \bibinfo{journal}{The Annals of Statistics} \bibinfo{volume}{33},
  \bibinfo{pages}{2873--2903}.
\bibitem[{Yuan and Cai(2010)}]{YuanCai10}
\bibinfo{author}{Yuan, M.}, \bibinfo{author}{Cai, T.T.}, \bibinfo{year}{2010}.
\newblock \bibinfo{title}{A reproducing kernel hilbert space approach to
  functional linear regression}.
\newblock \bibinfo{journal}{The Annals of Statistics} \bibinfo{volume}{38},
  \bibinfo{pages}{3412--3444}.

\end{thebibliography}

\end{document}